\documentclass[useAMS,usenatbib]{mn2e}
\usepackage{graphicx}

\title[The electron temperatures]
      {On the electron temperatures in high-metallicity HII regions}
      
\author[L.S.Pilyugin, L.Mattsson, J.M.V\'{\i}lchez,  and B.Cedr\'es
]
       {L.S.~Pilyugin$^{1}$
        and L.~Mattsson$^2$ 
        and J.M.~V\'{\i}lchez$^{3}$
         and B.~Cedr\'es$^{3}$ \\
     $^{1}$ Main Astronomical Observatory
            of National Academy of Sciences of Ukraine,
            27 Zabolotnogo str., 03680 Kiev, Ukraine \\
     $^{2}$ Department of Physics and Astronomy, Uppsala University,
            Uppsala, Sweden \\
     $^{3}$ Instituto de Astrof\'{\i}sica de Andaluc\'{\i}a,
            CSIC, Apdo, 3004, 18080 Granada, Spain \\
             }

\date{Accepted 2009 June 3. Received 2009 June 1; 
in original form 2009 May 12}

\begin{document}

\maketitle

\begin{abstract} 
The electron temperatures of high-metallicity (12+log(O/H) $> 8.2$) H\,{\sc ii} 
regions have been studied. The empirical ff relations which express the 
nebular-to-auroral [O\,{\sc iii}] line ratio $Q_{\rm 3,O}$ (as well as the 
nebular-to-auroral [O\,{\sc ii}] line ratio $Q_{\rm 2,O}$, and the 
nebular-to-auroral [N\,{\sc ii}] line ratio $Q_{\rm 2,N}$) in terms of the 
nebular R$_3$ and R$_2$ line fluxes in spectra of high-metallicity H\,{\sc ii} 
regions are derived, and the electron temperatures $t_{\rm 3,O}$, $t_{\rm 2,O}$, 
and $t_{\rm 2,N}$ in a number of extragalactic H\,{\sc ii} regions are also 
determined.  Furthermore, the $t_2$ -- $t_3$ diagram is discussed. It is found 
that there is a one-to-one correspondence between $t_2$ and $t_3$ electron 
temperatures for H\,{\sc ii} regions with a weak nebular R$_3$ lines (logR$_3$ 
$\la$ 0.5). The derived $t_{\rm 2,N}$ -- $t_{\rm 3,O}$ relation for these 
H\,{\sc ii} regions is similar to commonly used t$_2$ -- t$_3$ relations. 
The H\,{\sc ii} regions with a strong nebular R$_3$ line flux (logR$_3$ $\ga$ 
0.5) do not follow this relation. A discrepancy between $t_{\rm 2,N}$ and 
$t_{\rm 2,O}$ temperatures is found, being the $t_{\rm 2,N}$ temperatures 
systematically lower than $t_{\rm 2,O}$ ones. The differences are small at 
low electron temperatures and increases with increasing electron temperatures 
up to 10\% at $t = 1$. The uncertainties in the atomic data may be the cause 
of this discrepancy. 
\end{abstract} 
 
 
\section{Introduction} 
 
The two-zone model describing the electron temperatures within  
H\,{\sc ii} regions is commonly used for oxygen abundance determination 
\citep[see, e.g.,][]{campbell1986,garnett1992}. 
It is typically used when only the electron temperature $t_3$ is 
measured and the value of the other temperature $t_2$ is estimated using a 
$t_2$--$t_3$ relation. Since the form of the $t_2$--$t_3$ relation may be disputed  
\citep[and references therein]{pvt06,pilyugin07} the value of  
t$_2$ can be rather uncertain even if accurate measurements of $t_3$ are available.  
 
The majority of H\,{\sc ii} regions in spiral galaxies are low excitation objects  
\citep[e.g.]{zkh} and, consequently, the O$^+$ zone makes a significant (or even dominant)  
contribution to the total oxygen abundance. This shows that an accurate  
determination of $t_2$ is crucial for abundance determinations in H\,{\sc ii} regions of spiral galaxies. 
Furthermore, the nitrogen abundances in  H\,{\sc ii} regions   
are derived from a relation of the form ${\rm N/O} = f(t_2)$. This, again, underlines  that accurate  
determinations of $t_2$ is very important. 
 
The main objective of the present study is to find the electron temperatures in the 
zones of singly and doubly ionised oxygen, and a relation between them.  
The measured ratio $Q_{\rm 2,N}$ of nebular-to-auroral lines of singly ionised  
nitrogen is used to determine the electron temperature $t_{\rm 2,N}$ in the 
low-ionization zone of the H\,{\sc ii} regions, and the measured ratio $Q_{\rm 2,O}$  
of nebular-to-auroral lines of singly ionised oxygen is used to determine the  
electron temperature $t_{\rm 2,O}$. One may expect that $t_{\rm 2,N}$ and $t_{\rm 2,O}$  
are the same, or at least similar. The measured ratio $Q_{\rm 3,O}$ of nebular-to-auroral lines of  
doubly ionised oxygen is used to determine the electron temperature $t_{\rm 3,O}$ 
in the high-ionisation zone of the H\,{\sc ii} regions. 
 
A comparison between different electron temperatures reveals some problems.  
First of all, the precision of the measurements of the weak auroral lines  
[O\,{\sc ii}]$\lambda$7320+$\lambda$7330, [N\,{\sc ii}]$\lambda$5755, and   
[O\,{\sc iii}]$\lambda$4363 is usually low.  
Secondly, a comparison between different electron temperatures  
requires a set of H\,{\sc ii} regions where measurements of the auroral  
lines [O\,{\sc ii}]$\lambda$7320+$\lambda$7330, [N\,{\sc ii}]$\lambda$5755, and   
[O\,{\sc iii}]$\lambda$4363 are all available. Such measurements are quite scarce,  
but that problem may be tackled by constructing the corresponding ff relations.  
The ff relations allows us to estimate the $Q_{\rm 2,N}$, $Q_{\rm 2,O}$,  
and $Q_{\rm 3,O}$ values simultaneously (and, consequently, to determine  
$t_{\rm 2,N}$, $t_{\rm 2,O}$, and $t_{\rm 3,O}$ electron temperatures) in  
large samples of H\,{\sc ii} regions \citep{ff,pilyugin07}.  
                                                                              
This paper is organised as follows.  The adopted set of the atomic  
data used to convert the measured line ratios to the electron temperatures  
is discussed in Section 2. The ff relations for $Q_{\rm 3,O}$, $Q_{\rm 2,O}$  
and $Q_{\rm 2,N}$ are derived in Section 3.  
The relations between different temperatures are examined in Section 4. 
The results are discussed in Section 5  and Section 6 presents the conclusions. 
 
Throughout this paper, we will be using the following notations for the line 
fluxes: \\ 
R = [O\,{\sc iii}]$\lambda$4363 = $I_{{\rm [OIII]} \lambda 4363} /I_{{\rm H}\beta }$,  \\
R$_2$ = [O\,{\sc ii}]$\lambda$3727+$\lambda$3729  
      = $I_{[OII] \lambda 3727+ \lambda 3729} /I_{{\rm H}\beta }$,  \\
R$_3$ = [O\,{\sc iii}]$\lambda$4959+$\lambda$5007        = 
$I_{{\rm [OIII]} \lambda 4959+ \lambda 5007} /I_{{\rm H}\beta }$,  \\
R$_{23}$ = R$_2$ + R$_3$,  \\
\( \mbox{[O\,{\sc ii}]}\lambda 7325  
      = \mbox{[O\,{\sc ii}]} \lambda 7320+ \lambda 7330  
      =  I_{{\rm [OII]} \lambda 7320+ \lambda 7330} /I_{{\rm H}\beta } , \) \\ 
\( \mbox{[N\,{\sc ii}]}\lambda 5755  
      =  I_{{\rm [NII]} \lambda 5755} /I_{{\rm H}\beta } \), \\ 
N$_2$ = [N\,{\sc ii}]$\lambda$6548+$\lambda$6584  
      = $I_{{\rm [NII]} \lambda 6548+ \lambda 6584} /I_{{\rm H}\beta }$,  \\
S$_2$ = [S\,{\sc ii}]$\lambda$6717+$\lambda$6731  
      = $I_{{\rm [SII]} \lambda 6717+ \lambda 6731} /I_{{\rm H}\beta }$. \\ 
With these definitions,  the excitation parameter P can be expressed as:   \\
$P =$ R$_3$/(R$_2$+R$_3$), \\           
and the temperature indicators $Q_{2,O}$, 
$Q_{\rm 3, O}$ and $Q_{\rm 2,N}$ can be expressed as:  \\ 
$Q_{\rm 2, O}$ = R$_2$/[O\,{\sc ii}]$\lambda$7325,  \\  
$Q_{\rm 2, N}$ = N$_2$/[N\,{\sc ii}]$\lambda$5755,  \\ 
$Q_{\rm 3,O}$ = R$_3$/R. 
 
\section{Determination of electron temperatures} 
 
To convert the values of the $Q_{\rm 3,O}$, $Q_{\rm 2,O}$, and $Q_{\rm 2,N}$  
to the electron temperatures $t_{\rm 3,O}$, $t_{\rm 2,O}$, and $t_{\rm 2,N}$,  
we have used the five-level-atom solution for ions O$^{++}$, O$^+$, and  N$^+$  
together with recent atomic data.  
The Einstein coefficients for spontaneous transitions A$_{jk}$   
for five low-lying levels for all ions above were taken from \cite{froese2004}.  
The energy levels were taken from \citet{edlen1985} for O$^{++}$,  
from \citet{wenaker1990} for O$^+$, and from  
\citet{galavis1997} for N$^+$.  
The effective cross-sections, or effective collision strengths, for the electron  
impact $\Omega$$_{jk}$ were taken from \citet{aggarwal1999} for O$^{++}$,  
from \citet{pradhan2006} for O$^{+}$, and from \citet{hudson2005} for N$^{+}$.  
The effective cross-sections are continuous functions of temperatures,  
as tabulated by \citet{aggarwal1999,pradhan2006,hudson2005}  
at a fixed temperatures. The effective cross  
sections for a given electron temperature are derived from two-order  
polynomial fits to the data (from the studies cited above) as a function  
of temperature.  
 
In the low density regime ($n_e$ $<$ 500 cm$^{-3}$), the following simple  
expressions provide good approximations of the numerical results. 
For the $t_{\rm 3,O}$ -- $Q_{\rm 3,O}$ relation  
\begin{equation} 
t_{\rm 3,O}   = \frac{1.46}{\log Q_{\rm 3,O} + C_t}, 
\label{equation:t3o}    
\end{equation} 
where 
\begin{equation} 
C_t = - 0.88 - 0.17\log t_{\rm 3,O} + 0.030 t_{\rm 3,O},  
\end{equation} 
and for $t_{\rm 2,O}$ -- $Q_{\rm 2,O}$, 
\begin{equation} 
t_{\rm 2,O}   =  
 {0.96 \over \log Q_{\rm 2,O} + C_t}, 
\label{equation:t2o}    
\end{equation} 
where 
\begin{equation} 
C_t = - 0.86 - 0.38 \log t_{\rm 2,O}  
        +  0.053 t_{\rm 2,O} + \log(1+14.9 x_{\rm 2,O}),  
\end{equation} 
and 
\begin{equation} 
x_{\rm 2,O} = 10^{-4}\,n_e\, t_{\rm 2,O}^{-1/2} . 
\label{equation:x2o} 
\end{equation} 
Similarly, for the $t_{\rm 2,N}$ -- $Q_{\rm 2,N}$ relation  
\begin{equation} 
t_{\rm 2,N}   =  
 {1.12 \over \log Q_{\rm 2,N} + C_t},  
\label{equation:t2n}    
\end{equation} 
where  
\begin{equation} 
C_t =  - 0.89 - 0.19 \log t_{\rm 2,N}  
        +  0.032 t_{\rm 2,N} + \log(1+0.26 x_{\rm 2,N}),  
\end{equation} 
and 
\begin{equation} 
x_{\rm 2,N} = 10^{-4}\,n_e\, t_{\rm 2,N}^{-1/2} . 
\label{equation:x2n} 
\end{equation} 
These equations can be used instead of computing the electron 
temperature in low-density H\,{\sc ii} regions numerically.  
 
\section{ff relations} 
 
It has been argued that the excitation parameter $P$, 
combined with the measured  
fluxes of the strong nebular oxygen lines, can be used to estimate the  
physical conditions in the H\,{\sc ii} region \citep{lcal,hcal,vybor}. 
One may also say that the electron temperature in an H\,{\sc ii}  
region may be estimated using the nebular oxygen line fluxes only,  
since the excitation parameter $P$ in itself is defined in terms of the nebular oxygen  
line fluxes, i.e., the diagnostic line ratios can be expressed in terms  
of just the fluxes of the oxygen nebular lines. It has been found empirically  
that there exist a relation (the ff relation) between auroral  
[O\,{\sc iii}]$\lambda$4363 and nebular oxygen line fluxes in spectra of  
H\,{\sc ii} regions \citep{ff,pilyuginetal06}.  
It has been also shown that there exist a relation between $Q_{\rm 2,N}$  
and nebular oxygen-line fluxes in spectra of H\,{\sc ii} regions \citep{pilyugin07}.  
Here, relations of this type are derived for $Q_{\rm 3,O}$, $Q_{\rm 2,O}$ and $Q_{\rm 2,N}$.

\subsection{A sample of the calibration data} 
 
The reliability of the derived ff relations depend on the quality of 
the calibration data, i.e., on the accuracy of the auroral 
[N\,{\sc ii}]$\lambda$5755, [O\,{\sc ii}]$\lambda$7325 and 
[O\,{\sc iii}]$\lambda$4363 line measurements in the H\,{\sc ii} regions, 
which are used in deriving the relations. 
We have compiled a sample of the most reliable spectrophotometric measurements 
of H\,{\sc ii} regions from the literature in the following way. 
First of all, we have only considered recent (published after 2000) observations and
line measurements in the spectra of H\,{\sc ii} regions. 
Then, we selected the most reliable measurements based on the following criteria. 
The main requirement is that the uncertainty in the measured 
electron temperature $t_{\rm 3,O}$ (or $t_{\rm 2,O}$, or $t_{\rm 2,N}$) 
is less than around 10\%. The estimated uncertainties given in the original publications 
were used. 
Futhermore, if the measurements of the electron temperatures for two or more 
ions (e.g. O$^{++}$ and S$^{++}$) were available for the H\,{\sc ii} region  
then the agreement (or disagreement) between these temperatures was used as an 
additional criterion: the difference between 
electron temperatures derived for O$^{++}$ and S$^{++}$ should not be in 
excess of 1000 K. 

We believe that our criteria allow us to pick out the spectra of H\,{\sc ii} regions 
with reliable measurements of at least one auroral line 
([N\,{\sc ii}]$\lambda$5755, [O\,{\sc ii}]$\lambda$7325 and 
[O\,{\sc iii}]$\lambda$4363). 
Our final list consists of 76 spectra of H\,{\sc ii} regions taken from 
\citet{guseva00,luridianaetal02,vermeijetal02,kennicuttetal03,leeetal03,
tsamisetal03,peimbert03,vilchez03,bresolinetal04,izotovthuan04,  
garnettetal04,leeskillman04,leeetal04,bresolinetal05,thuan05,   
crockett06,bresolin07}.  
The measured [O\,{\sc ii}]$\lambda$3727+$\lambda$3729, 
[O\,{\sc iii}]$\lambda$4363, [O\,{\sc iii}]$\lambda$4959+$\lambda$5007, 
[N\,{\sc ii}]$\lambda$5755, [N\,{\sc ii}]$\lambda$6548+$\lambda$6584, 
[S\,{\sc ii}]$\lambda$6717+$\lambda$6731, [O\,{\sc ii}]$\lambda$7320+$\lambda$7330 
line intensities are listed in Table \ref{table:lines} which is available online. 
In cases where the $I_{{\rm [O\,III]}\lambda 4959}$ line is not available, 
the R$_3$ is obtained from the relation 
R$_3$ = 1.33 $\times$ $I_{{\rm [O\,III]}\lambda 5007}$/$I_{{\rm H}\beta }$.  
Similarly, if the $I_{{\rm [N\,II]}\lambda 6548}$ line is not available, 
the N$_2$ is obtained from the relation 
N$_2$ = 1.33 $\times$ $I_{{\rm [N\,II]}\lambda 6584}$/$I_{{\rm H}\beta }$.  
It should be noted that if the accuracy of any measured auroral line is 
expected to be low, the line intensity is set to zero in Table \ref{table:lines}. 
 
After this investigation has been completed, a
new spectrophotometric measurements for H\,{\sc ii} regions have been 
published by \citet{bresolinetal09,estebanetal09}. Those data are used to 
test the reliability of the relations derived in the present paper.

\subsection{The ff relation for $Q_{\rm 2,N}$} 

\begin{figure} 
\resizebox{1.00\hsize}{!}{\includegraphics[angle=000]{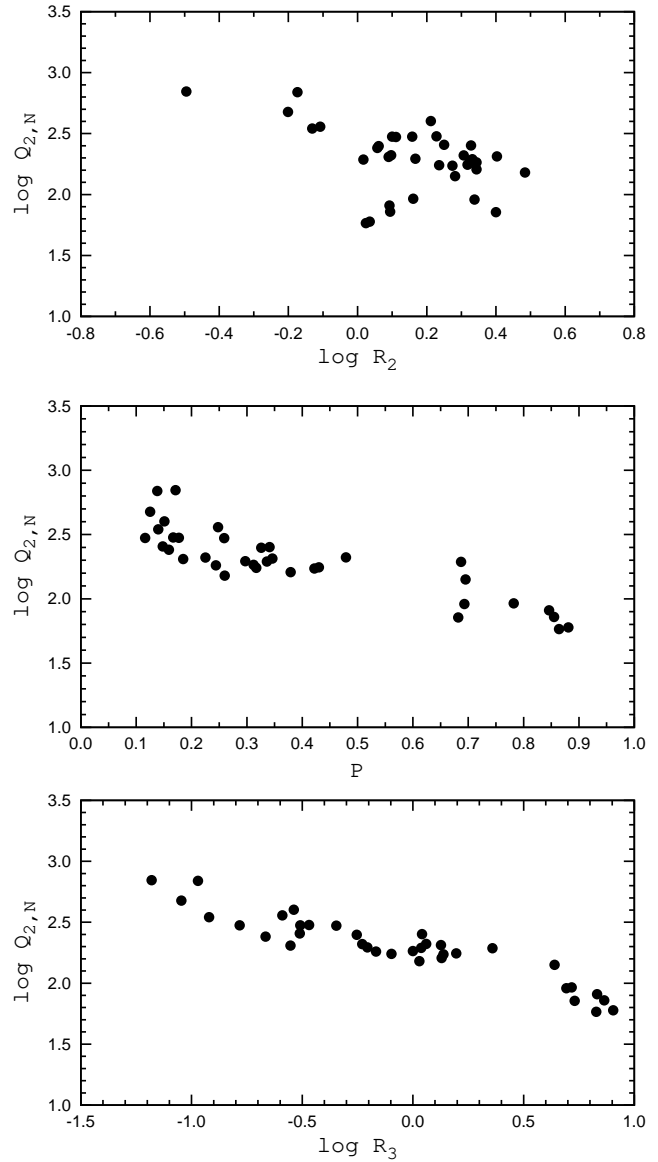}} 
\caption{ 
The electron temperature indicator $Q_{\rm 2,N}$ as a function of  
the R$_2$ line flux ({\it top panel}), excitation parameter ({\it middle panel}) 
and R$_3$ line flux ({\it bottom panel}).   
} 
\label{figure:q2n} 
\end{figure} 
 
\begin{figure} 
\resizebox{1.00\hsize}{!}{\includegraphics[angle=000]{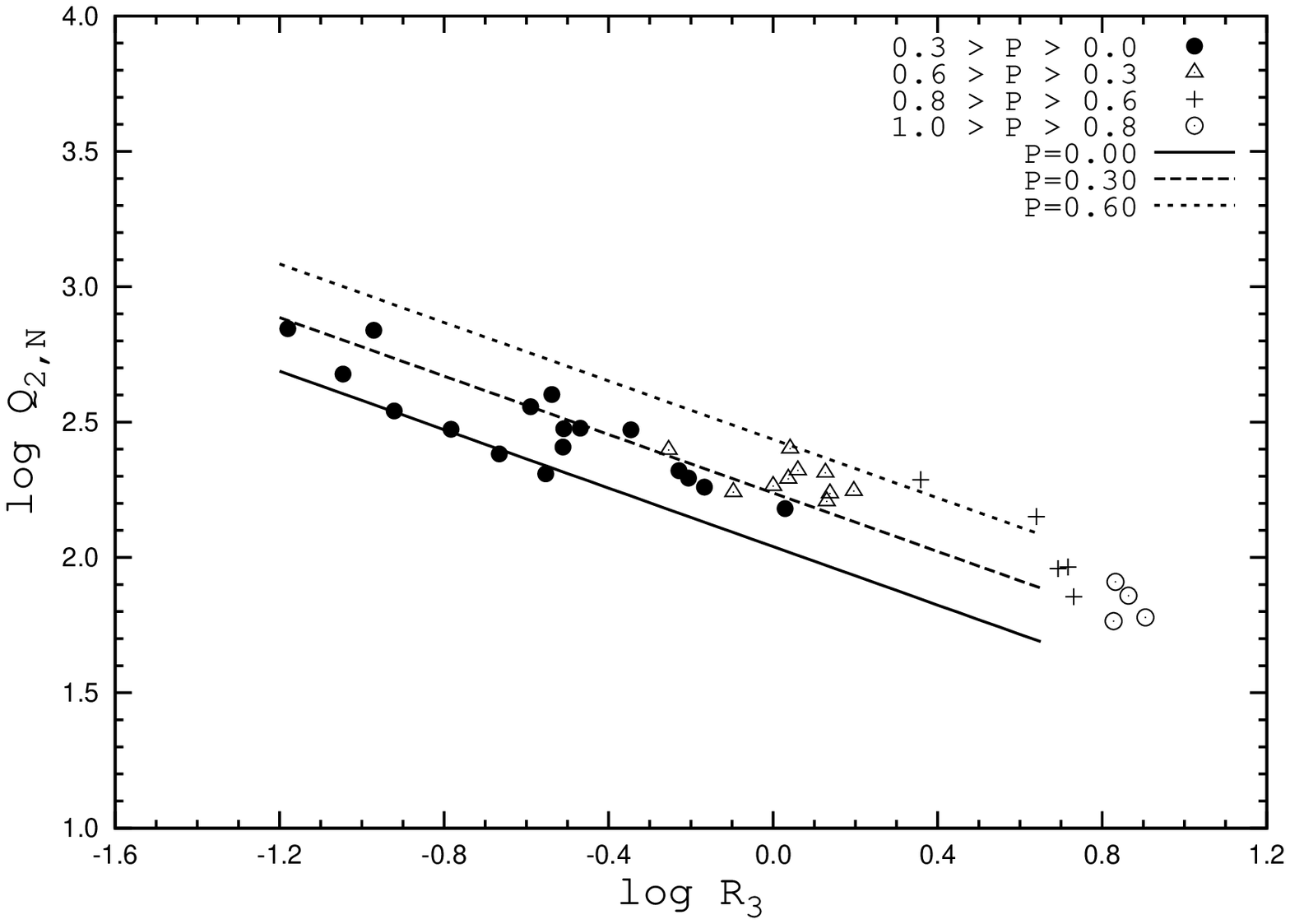}} 
\caption{ 
The electron-temperature indicator $Q_{\rm 2,N}$ as function of the R$_3$ 
line flux.  
The family of $Q_{\rm 2,N}=f($R$_3,P)$ curves for different values of the  
excitation parameter is superimposed on the observational data. 
} 
\label{figure:r3-q2n} 
\end{figure} 
 
\begin{figure} 
\resizebox{1.00\hsize}{!}{\includegraphics[angle=000]{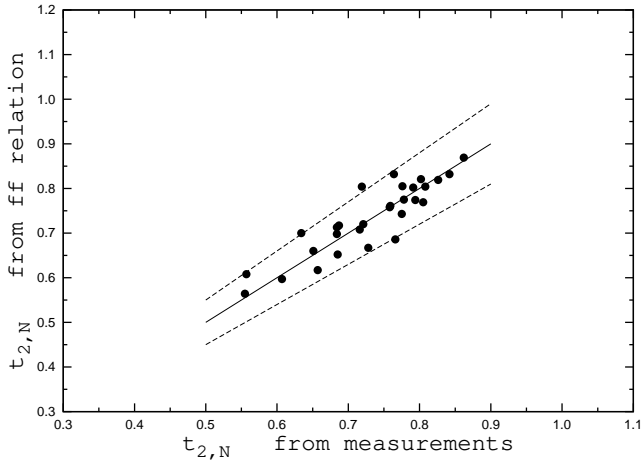}} 
\caption{ 
The electron temperature $t_{\rm 2,N}$ estimated from the ff relations  
against the measured electron temperature $t_{\rm 2,N}$.  
The filled circles are the calibration H\,{\sc ii} regions. 
The solid line shows the case of equal values and the dashed lines show  
$\pm$10\% deviations. 
} 
\label{figure:t2n} 
\end{figure}

Fig.~\ref{figure:q2n} shows the electron temperature indicator $Q_{\rm 2,N}$  
as a function of the R$_2$ line flux (top panel), excitation parameter $P$  
(middle panel) and the R$_3$ line flux (bottom panel) for the 
sampled spectra of H\,{\sc ii} regions, where the auroral nitrogen line  
[N\,{\sc ii}]$\lambda$5755 is available. 
It is evident from Fig.~\ref{figure:q2n} that the $Q_{\rm 2,N}$ values  
show a stronger correlation with the R$_3$ line flux than with the R$_2$ line 
flux.  H\,{\sc ii} regions with different values of the excitation  
parameter are shown in Fig.~\ref{figure:r3-q2n} by different symbols; 
0.3 $> P >$ 0.0 by filled circles, 0.6 $> P >$ 0.3 by triangles, 
0.8 $> P >$ 0.6 by plus signs, 1.0 $> P >$ 0.8 by open circles. 
Fig.~\ref{figure:r3-q2n} shows that the $Q_{\rm 2,N}$ values  
correlate both with the R$_3$ line flux and the excitation parameter $P$.  
It has been found that an expression of the simple form  
\begin{equation} 
\log Q_{2,N} = a_1 + a_2 \, P + a_3 \, \log R_3  
\label{equation:qtype}    
\end{equation} 
reproduces the calibration data quite well. 
 
The coefficients in Eq.(\ref{equation:qtype}) are found using our sample  
of calibration H\,{\sc ii} regions as described in \citet{ff,pilyuginthuan07}.   
We obtain  
\begin{equation} 
\log Q_{2,N} = 2.04 + 0.66 \, P - 0.56 \, \log R_3  , \;\;\ \log R_3 < 0.5 .  
\label{equation:q2nl}    
\end{equation} 
The coefficients in Eq.(\ref{equation:q2nl}) are found using calibration 
data with relatively weak R$_3$ lines $(\log$ R$_3$ $< 0.5)$.  
The calibration data with strong R$_3$ lines (logR$_3$ $\ga$ 0.5)  
do not follow this relation. A gap in the calibration data with logR$_3$ 
around 0.5 prevents us to establish a solid value of the limit of the 
applicability of the derived relation. This (conservative) value was estimated 
on the base of other diagram (see Section 5).
It is possible that another set of coefficients  
should be derived for these objects, or a more complex functional form should be  
used for the ff relation in order to reproduce the measurements over the whole 
interval of R$_3$ fluxes. This would require more  
calibration data points (in particular those with strong R$_3$ lines). 
The family of $Q_{\rm 2,N}=f($R$_3,P)$ curves for different values of the  
excitation parameter is shown in Fig.~\ref{figure:r3-q2n}, superimposed on the 
observational data.  The curve with $P=0.0$ is shown by the  
solid line, the long-dashed line shows the $P=0.30$ case, and  
the short-dashed line shows the $P=0.6$ case.  
 
Fig.~\ref{figure:t2n} shows the electron temperature $t_{\rm 2,N}$ estimated  
from the derived ff relation and plotted against the measured electron  
temperature $t_{\rm 2,N}$. The filled circles show the H\,{\sc ii}  
regions used for calibration. The solid line shows the case of equal values 
and the dashed lines show  $\pm$10\% deviations.

\subsection{The ff relation for $Q_{\rm 2,O}$} 
 
\begin{figure} 
\resizebox{1.00\hsize}{!}{\includegraphics[angle=000]{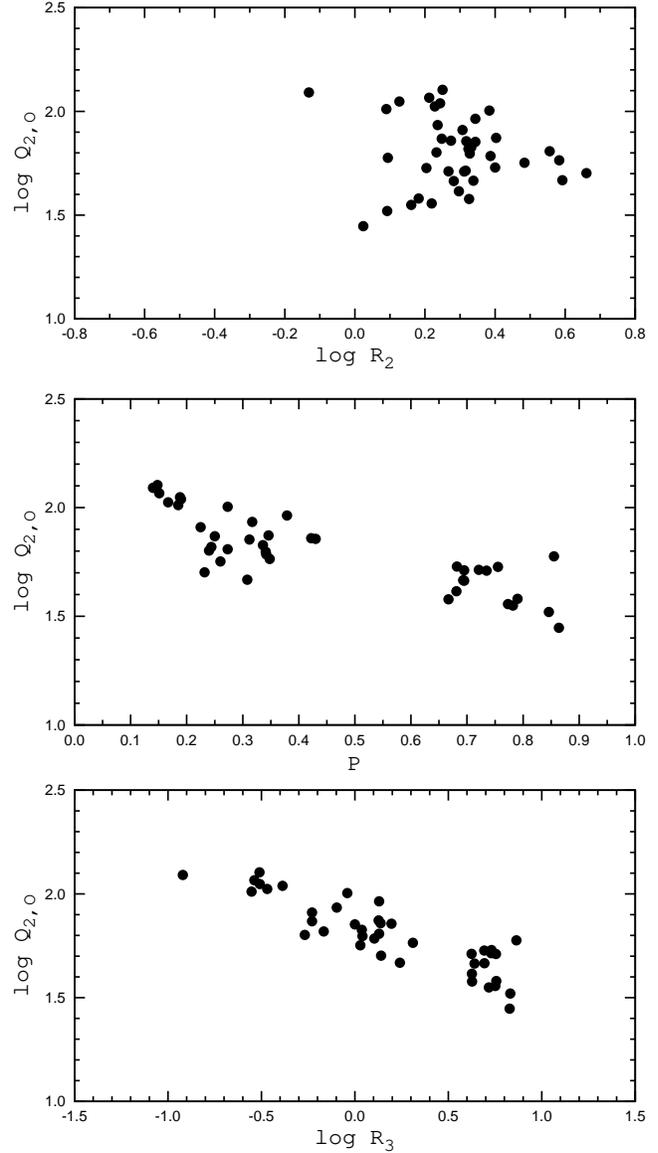}} 
\caption{ 
The electron temperature indicator $Q_{\rm 2,O}$ as a function of  
flux the R$_2$ line flux ({\it top panel}), excitation parameter ({\it middle 
panel}) and R$_3$ line flux ({\it bottom panel}).   
} 
\label{figure:q2o} 
\end{figure} 
 
\begin{figure} 
\resizebox{1.00\hsize}{!}{\includegraphics[angle=000]{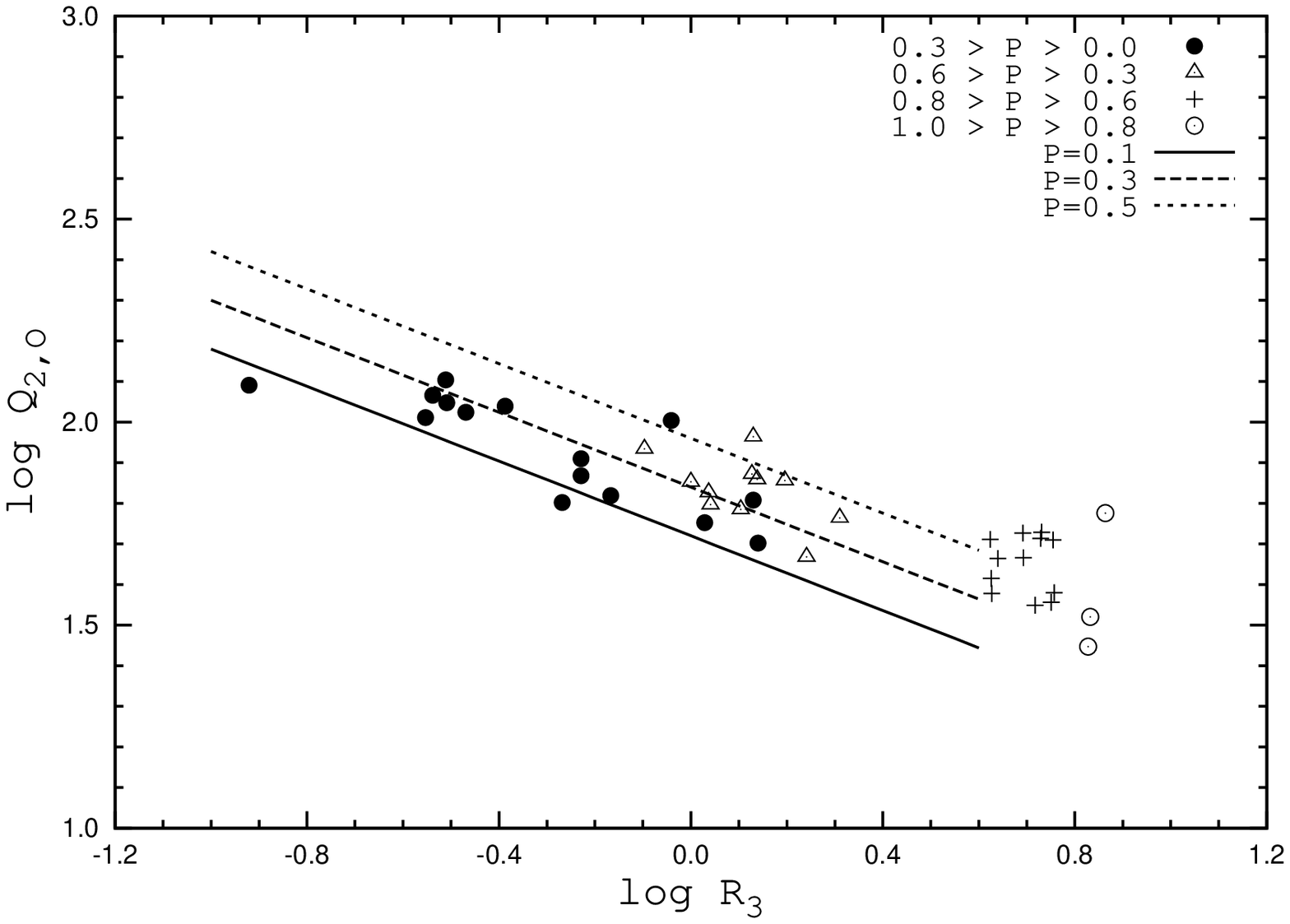}} 
\caption{ 
The electron temperature indicator $Q_{\rm 2,O}$ against the R$_3$ line flux.  
The family of $Q_{\rm 2,N} =f($R$_3,P)$ curves for different values of the  
excitation parameter is superimposed on the observational data.  
} 
\label{figure:r3-q2o} 
\end{figure} 
 
\begin{figure} 
\resizebox{1.00\hsize}{!}{\includegraphics[angle=000]{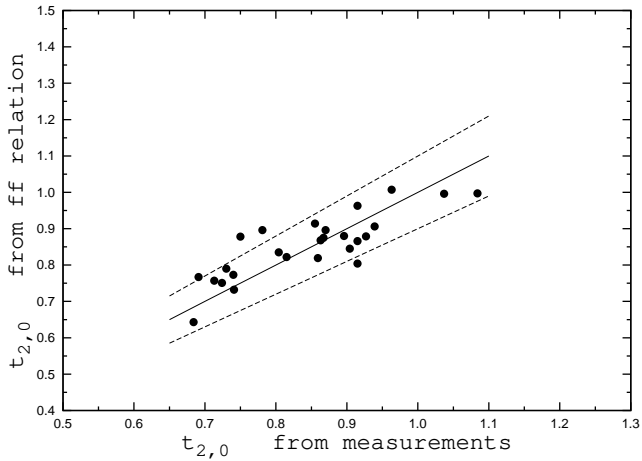}} 
\caption{ 
The electron temperature $t_{\rm 2,O}$ estimated from the ff relations  
against the measured electron temperature $t_{\rm 2,O}$.  
The filled circles show the calibration H\,{\sc ii} regions. 
The solid line shows the case of equal values and the dashed lines show $\pm$10\% deviations. 
} 
\label{figure:t2o} 
\end{figure} 
 
The sub-sample of H\,{\sc ii} regions, where the auroral oxygen line 
[O\,{\sc ii}]$\lambda$7325 is available, has been analysed in a way  
similar to that of the sub-sample considered in the previous section 
(see Figs.~\ref{figure:q2o}--\ref{figure:t2o}). 
We find the following ff relation for [O\,{\sc ii}] lines in the low-R$_3$ range,  
($\log$~R$_3 < 0.5$)  
\begin{equation} 
\log Q_{2,O} = 1.66 + 0.60 \, P - 0.46 \, \log R_3  , \;\;\ \log R_3 < 0.5 .  
\label{equation:q2ol}    
\end{equation} 
Again, the calibration data with strong R$_3$ lines  
(logR$_3 \ga 0.5$) do not follow this relation.

\subsection{The ff relation for $Q_{\rm 3,O}$} 

\begin{figure} 
\resizebox{1.00\hsize}{!}{\includegraphics[angle=000]{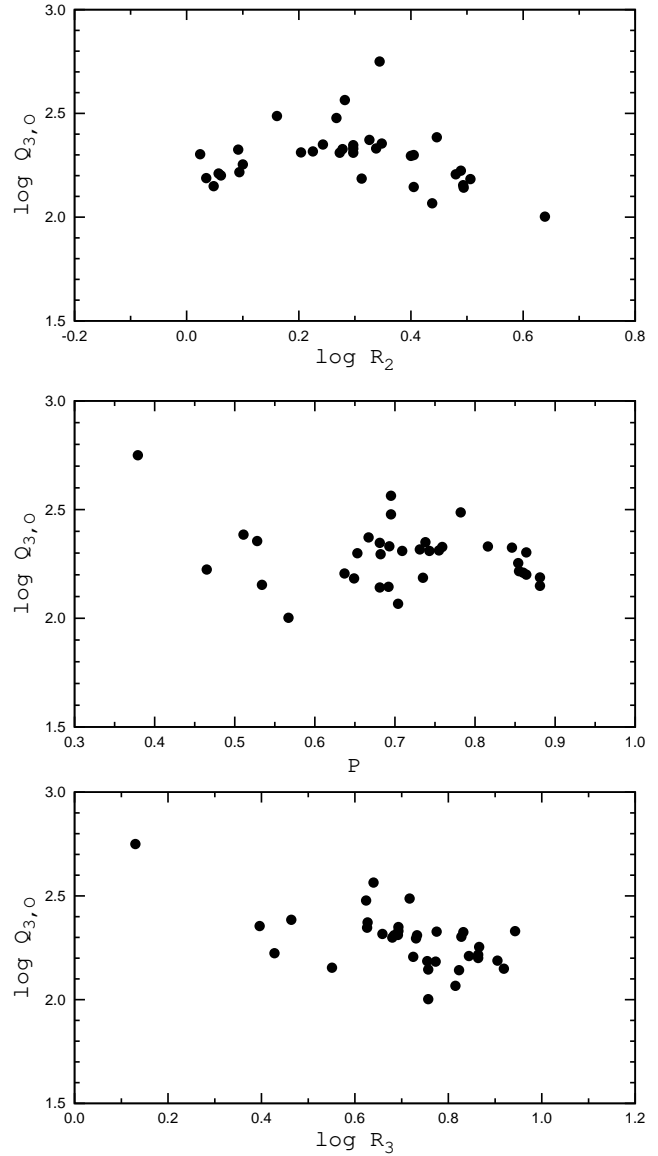}} 
\caption{ 
The electron temperature indicator $Q_{\rm 3,O}$ as a function of  
the R$_2$ line flux ({\it top panel}), excitation parameter ({\it middle panel}) 
and R$_3$ line flux ({\it bottom panel}).   
} 
\label{figure:q3o} 
\end{figure} 
 
\begin{figure} 
\resizebox{1.00\hsize}{!}{\includegraphics[angle=000]{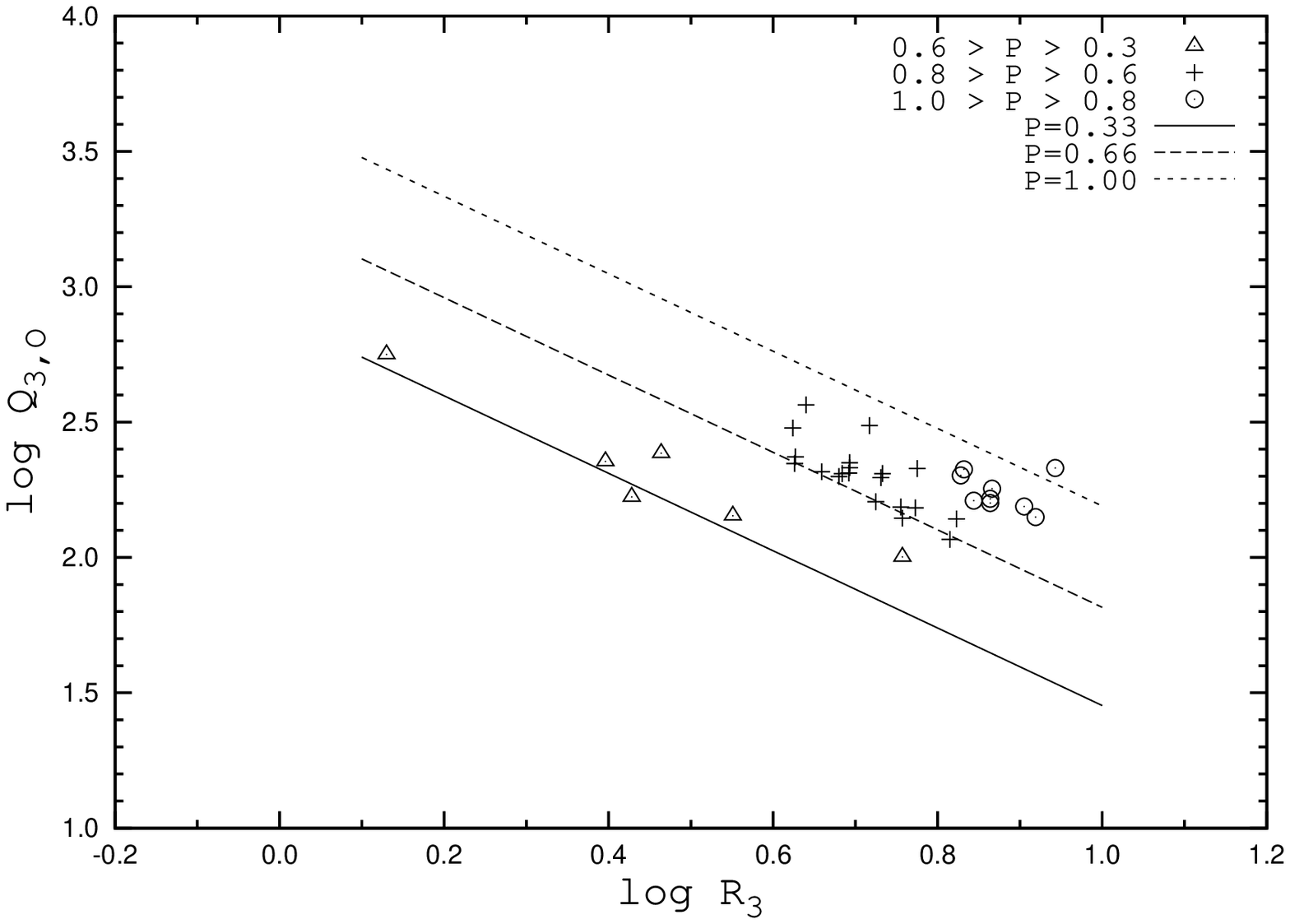}} 
\caption{ 
The electron temperature indicator $Q_{\rm 3,O}$ against flux in the  
line R$_3$.  
The family of $Q_{\rm 2,N}=f($R$_3,P)$ curves for different values of the  
excitation parameter is superimposed on the observational data. 
} 
\label{figure:r3-q3o} 
\end{figure} 
 
\begin{figure} 
\resizebox{1.00\hsize}{!}{\includegraphics[angle=000]{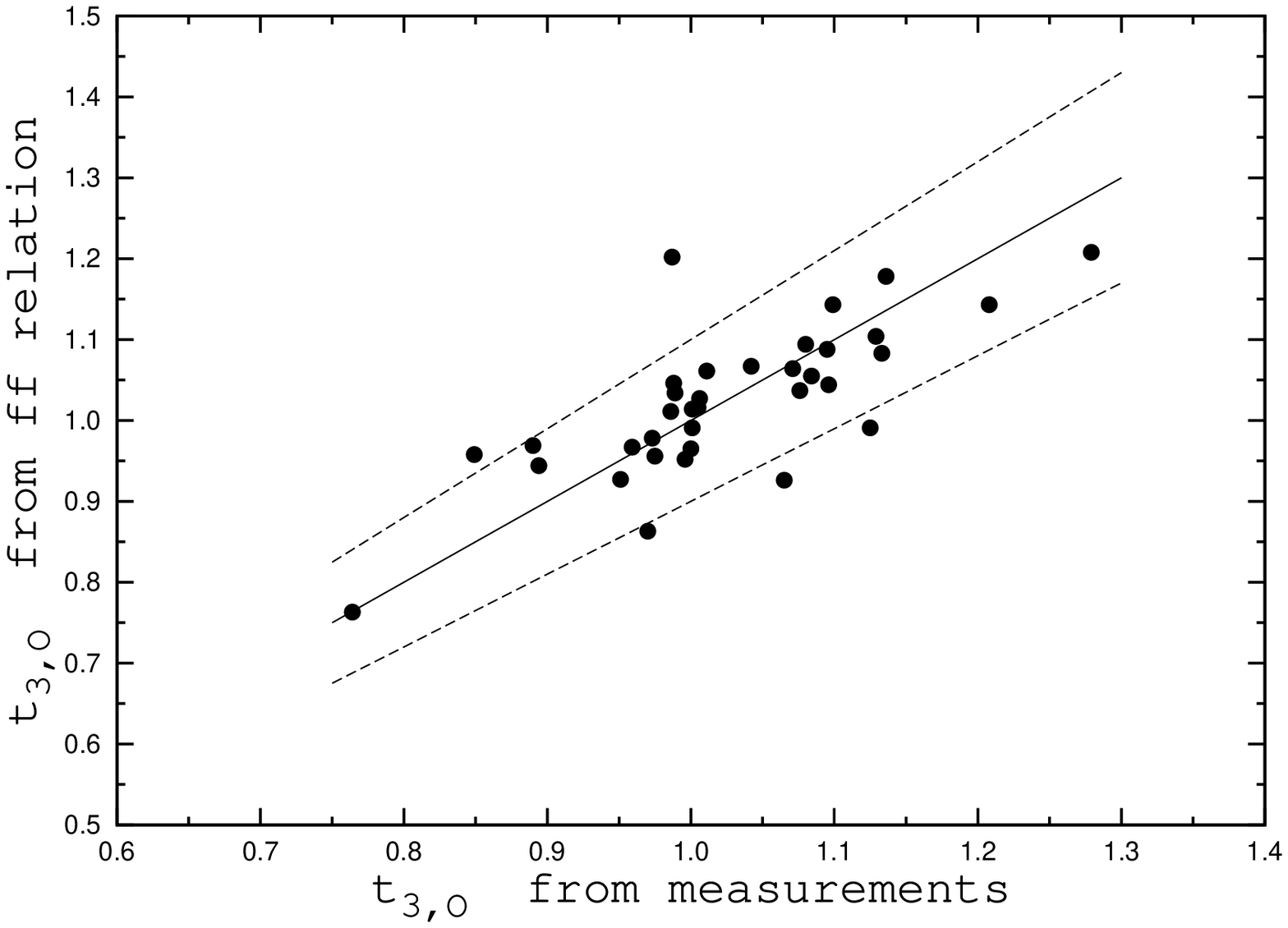}} 
\caption{ 
The electron temperature $t_{\rm 3,O}$ estimated from the ff relations  
against the measured electron temperature $t_{\rm 3,O}$.  
The filled circles show the calibration H\,{\sc ii} regions. 
The solid line shows the case of equal values and the dotted lines show $\pm$10\% deviations. 
} 
\label{figure:t3o} 
\end{figure}

An analysis of the sub-sample of H\,{\sc ii} regions, where the auroral oxygen 
line [O\,{\sc iii}]$\lambda$4363 is available 
(see Figs.~\ref{figure:q3o}--\ref{figure:t3o})  
results in the following ff relation for [O\,{\sc iii}] lines 
\begin{equation} 
\log Q_{3,O} = 2.53 + 1.08 \, P - 1.42 \, \log R_3. 
\label{equation:q3o}    
\end{equation} 
In this case, the calibration data follow the derived relation over 
the whole range of R$_3$ values.

\section{Relations between electron temperatures}  
 
\subsection{Comparison between $t_{2,N}$ and $t_{2,O}$ temperatures} 
 
\begin{figure} 
\resizebox{1.00\hsize}{!}{\includegraphics[angle=000]{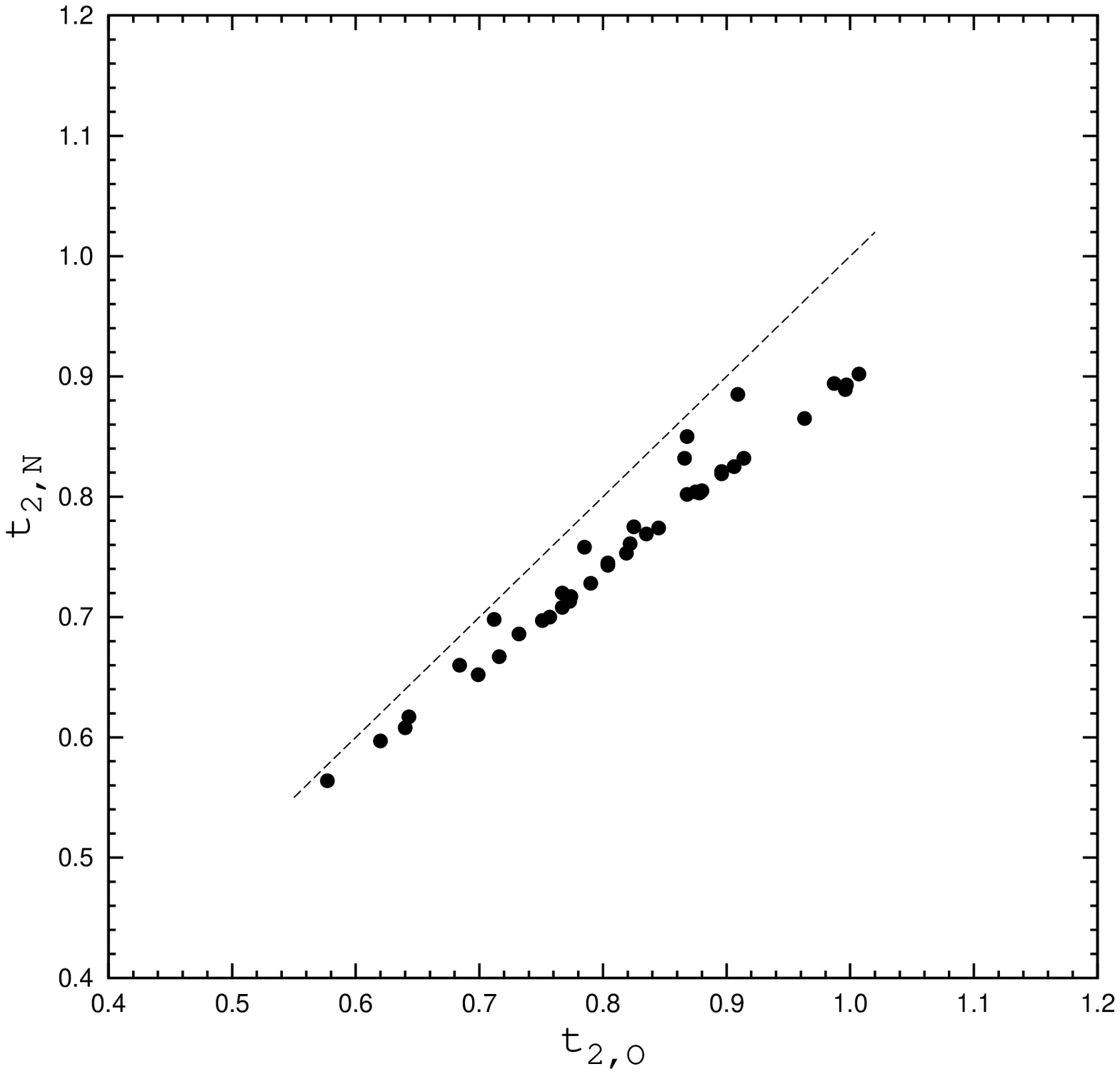}} 
\caption{ 
Electron temperature $t_{\rm 2,N}$ against electron temperature $t_{\rm 2,O}$.  
The filled circles show $t_{\rm 2,N}$ and $t_{\rm 2,O}$ in our samples of  
extragalactic H\,{\sc ii} regions with logR$_3$ $<$ 0.5 estimated using the ff 
relations.  The dashed line shows the case of equal values.  
} 
\label{figure:t2t2} 
\end{figure}

\begin{figure} 
\resizebox{1.00\hsize}{!}{\includegraphics[angle=000]{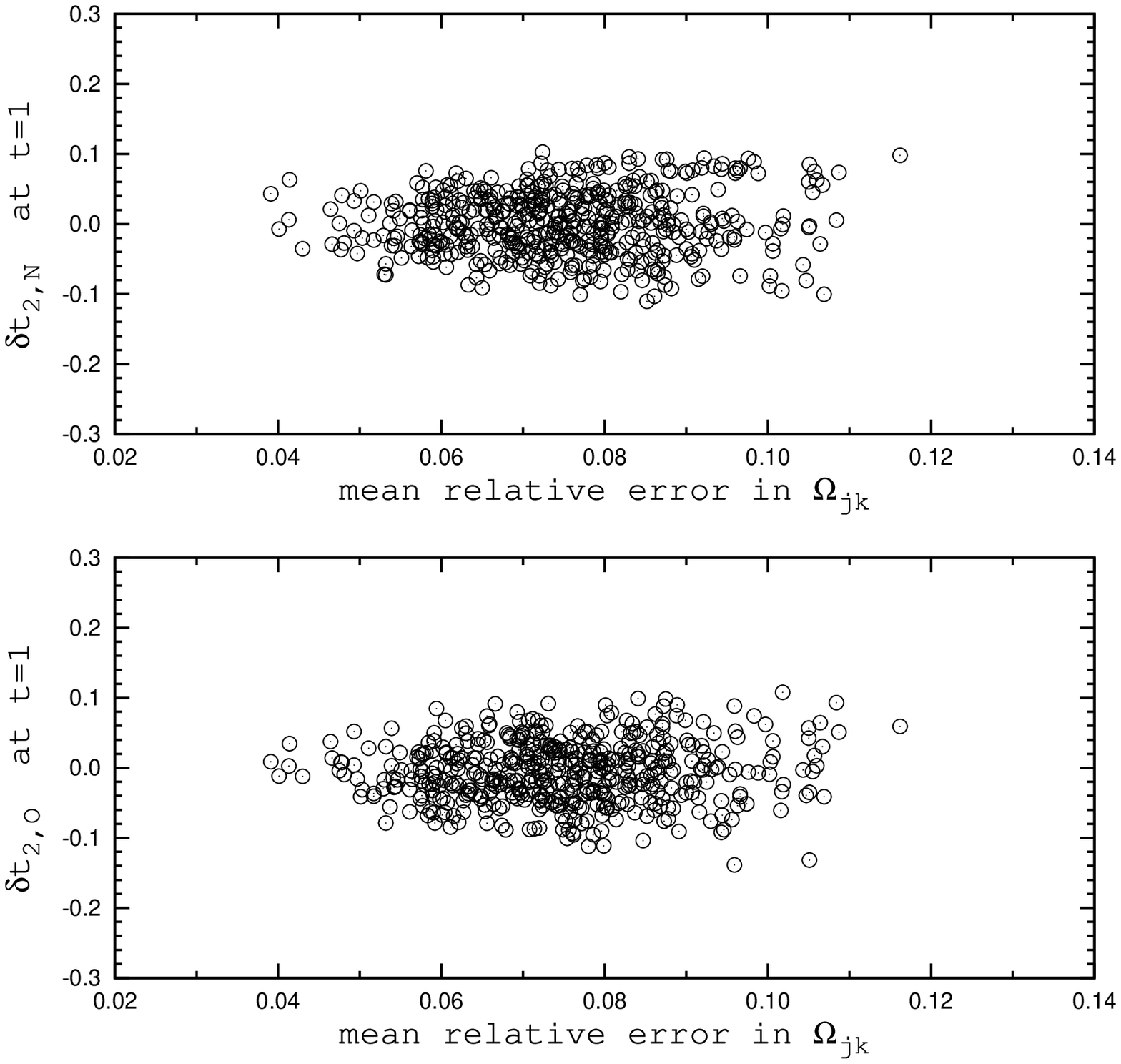}} 
\caption{ 
{\it Upper panel:} the relative difference (at $t=1$) between the electron 
temperature 
t$^*_{\rm 2,N}$ derived with  a random error component added to the 
 effective cross-sections for electron impact $\Omega$$_{jk}$ and 
the electron temperature $t_{\rm 2,N}$  derived with ``standard'' atomic data 
as a function of the average of the absolute values of the relative random error 
components. 
{\it Lower panel:} same as upper panel, but for the electron temperature 
$t_{\rm 2,O}$.  
} 
\label{figure:et2n} 
\end{figure} 
 
Fig.~\ref{figure:t2t2} shows the $t_{\rm 2,N}$ -- $t_{\rm 2,O}$ diagram.  
The values of $t_{\rm 2,N}$ and $t_{\rm 2,O}$ for our sample of  
extragalactic H\,{\sc ii} regions with $\log$ R$_{3} < 0.5$, estimated using 
the ff relations, are shown by filled circles (the ff-based temperatures  
are used because measurements of both $t_{\rm 2,N}$ and  
$t_{\rm 2,O}$ temperatures are not available). The dashed line shows the case of  
equal values. It is evident from Fig.~\ref{figure:t2t2} that the  
$t_{\rm 2,N}$ temperatures are systematically lower than $t_{\rm 2,O}$  
temperatures. At low electron temperatures the discrepancy is small, but it  
increases with increasing electron temperature. The difference is around 
10\% at $t_2 = 1.0$. We find four possible causes for this discrepancy:  
 
\begin{itemize} 
\item[(1)] The conversion of the strong line fluxes into the temperature  
indicators $Q_{\rm 2,N}$ and $Q_{\rm 2,O}$ involve systematic errors, i.e.,  
the ff relation for [N\,{\sc ii}] lines or/and the ff relation for  
[O\,{\sc ii}] lines is/are too crude and involves a systematic error.  
If this is the case, low precision of the line measurements in  
the spectra of the calibration H\,{\sc ii} regions can be the cause of the  
discrepancy between $t_{\rm 2,N}$ and $t_{\rm 2,O}$.        
 
\item[(2)] The conversion of the temperature indicators $Q_{\rm 2,N}$ or/and  
$Q_{\rm 2,O}$ to the electron temperatures $t_{\rm 2,N}$ or/and to   
$t_{\rm 2,O}$ involves systematic errors.  
If this is the case, low precision of the atomic data can be the cause of the  
the discrepancy between $t_{\rm 2,N}$ and $t_{\rm 2,O}$.  
 
\item[(3)] The value $Q_{2,O}$ and/or $Q_{2,N}$ is/are bad indicators of the  
electron temperatures and do not reflect the true electron temperature in an 
H\,{\sc ii} region.  
 
\item[(4)] The the [N\,{\sc ii}] lines are sometimes blended with the H$\alpha$  
lines. In such cases it is possible that the electron temperatures $t_{\rm 2,N}$ 
are affected by the errors occurring when disentangling these two lines. It is 
quite likely that this is mostly a problem for objects where emission lines are  
strong, but it is difficult to quantify the effects. Hence, we here only mention this 
possibility and do not analyse it in any further detail. 
\end{itemize} 
We have also considered the effects of absorption from an underlying stellar population. 
But since absorption effects are pronounced mostly in the Balmer lines, the impact on, e.g., 
R$_2$ and R$_3$ is negligible and cannot be the cause of the discrepancy described above. 
 
Uncertainties in the electron temperatures $t_{\rm 3,O}$, $t_{\rm 2,O}$  
and $t_{\rm 2,N}$ caused by uncertainties in the atomic data have been  
considered by \citet{pilyugin09}. Following that work, we like to point out  
a few critical issues, regarding such uncertainties. 
To convert the temperature indicator $Q_{\rm 2,N}$ into the electron temperature  
$t_{\rm 2,N}$, we have used the five-level-atom solution for N$^+$  
with the Einstein coefficients for spontaneous transitions A$_{jk}$ obtained 
by \cite{froese2004} and the effective cross-sections for electron  
impact $\Omega$$_{jk}$ from \citet{hudson2005}.  
It is expected that the best present effective collision strengths between the 
low-lying states have a typical uncertainty of around 10\% 
\citep{hudson2004,tayal2007}.
The general accuracy of the present 
probabilities is expected to be within 10\% \citep{galavis1997}. 
 
The uncertainty in $t_{\rm 2,N}$, caused by the  
uncertainty in the atomic data, can be estimated in the following way.  
The values of A$_{jk}$ from \cite{froese2004} and $\Omega$$_{jk}$ from  
\citet{hudson2005} have been considered as ``standard'' values. We then  
introduce a random relative error from -15\% to +15\% to every value of 
A$_{jk}$ and  $\Omega$$_{jk}$.  The electron  temperature derived from 
the five-level-atom solution with an random-error component added   
to the atomic data will be referred to as $t^*_{\rm 2,N}$.  
The $t^*_{\rm 2,N}$ values for a set of $Q_{\rm 2,N}$ values were computed with  
500 different random errors added to the atomic data. The 
relative error in the $t_{\rm 2,N}$ temperature  $\delta$$t_{\rm 2,N}$ = 
1.- $t^*_{\rm 2,N}$/$t_{\rm 2,N}$ caused by the uncertainties in the atomic 
data, increases with increasing of electron temperature.  
The top panel of  Fig.~\ref{figure:et2n} shows $\delta$$t_{\rm 2,N}$ 
at $t=1$ versus the mean of the absolute values of the relative random error 
component in $\Omega$$_{jk}$. 
The value of $\delta$$t_{\rm 2,N}$ is mainly governed by the 
uncertainties in the effective collision strengths for transitions from 
level 1 to levels 4 and 5.  
When the errors in $\Omega$$_{14}$ and $\Omega$$_{15}$ are large (and have 
opposite signs), the value of $\delta$$t_{\rm 2,N}$ is large even if 
the mean of the absolute values of the random error component in 
$\Omega$$_{jk}$ are small due to small errors in other $\Omega$$_{jk}$. 
On the other hand, when the errors in $\Omega$$_{14}$ and $\Omega$$_{15}$ are small 
then the value of $\delta$$t_{\rm 2,N}$ is small, even if 
the mean of the absolute values of the random error component in 
$\Omega$$_{jk}$ are large due to large errors in other $\Omega$$_{jk}$. 
 
To convert the temperature indicator $Q_{\rm 2,O}$ into the electron temperature  
value $t_{\rm 2,O}$, we have used the five-level-atom solution for O$^+$  
with the Einstein coefficients for spontaneous transitions A$_{jk}$    
obtained by \cite{froese2004} and the effective cross sections for electron  
impact $\Omega$$_{jk}$ from \citet{pradhan2006}.  
Again, the t$^*_{\rm 2,O}$ values for a set of $Q_{\rm 2,O}$ values were 
computed with  500 different random errors (from -15\% to +15\%) added to 
every value of A$_{jk}$ and  $\Omega$$_{jk}$.   The bottom panel of  
Fig.~\ref{figure:et2n} shows $\delta$$t_{\rm 2,O}$ 
at $t=1$ versus the mean of the absolute values of the random error 
component in $\Omega$$_{jk}$. 
The Fig.~\ref{figure:et2n} shows that the value of the error in the electron 
temperatures $t_{\rm 2,O}$ and $t_{\rm 2,O}$ caused by  
uncertainties in the atomic data is comparable with the obtained discrepancy 
between $t_{\rm 2,N}$ and $t_{\rm 2,O}$. 

We find that the errors in the electron temperatures $t_{\rm 2,N}$ and/or $t_{\rm 2,O}$  
caused by uncertainties in the atomic data can explain the  
discrepancy between the electron temperatures $t_{\rm 2,N}$ and  
$t_{\rm 2,O}$ that we obtain. However, it should be noted  that the errors caused by uncertainties in the  
ff relation for [O\,{\sc ii}] lines and/or the ff relation for  
[N\,{\sc ii}] lines can also make a contribution to the discrepancy  
between $t_{\rm 2,N}$ and $t_{\rm 2,O}$.  
More high-precision measurements of nitrogen and oxygen  
auroral lines are needed in order to find better ff relations and to elucidate the  
origin of the discrepancy between $t_{\rm 2,N}$ and $t_{\rm 2,O}$.  
 
\subsection{The relation between electron temperature in low- and  
high-ionisation zones} 
 
\begin{figure} 
\resizebox{1.00\hsize}{!}{\includegraphics[angle=000]{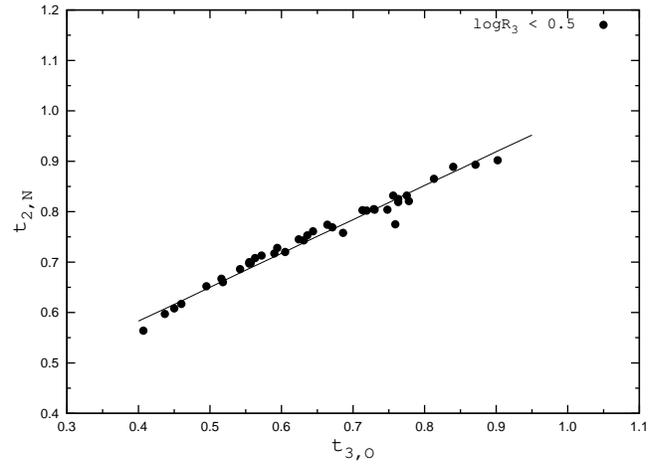}} 
\caption{ 
The $t_{\rm 2,N}$ -- $t_{\rm 3,O}$ diagram for H\,{\sc ii} regions 
with weak R$_3$ line fluxes ($\log$~R$_3 < 0.5$). 
The ff-based $t_{2,N}$ and $t_{3,O}$ temperatures are used. 
The solid line shows the derived $t_{\rm 2,N}$ -- $t_{\rm 3,O}$ relation. 
} 
\label{figure:t3o-t2n} 
\end{figure}

\begin{figure} 
\resizebox{1.00\hsize}{!}{\includegraphics[angle=000]{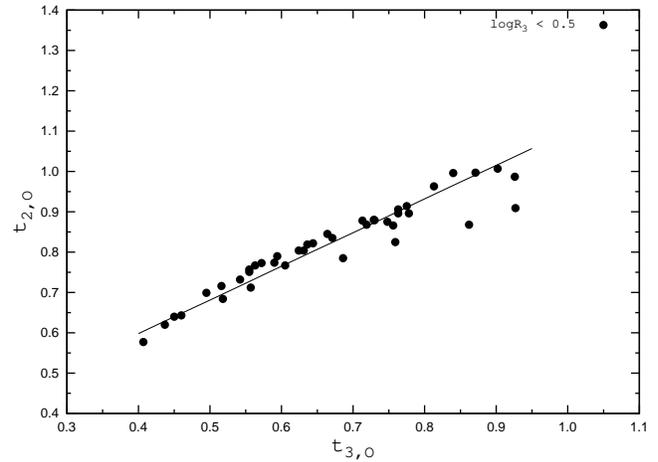}} 
\caption{The same as Fig.~\ref{figure:t3o-t2n} but for $t_{\rm 2,O}$ -- 
$t_{\rm 3,O}$ relation.
} 
\label{figure:t3o-t2o} 
\end{figure} 
 
Fig.~\ref{figure:t3o-t2n} shows the $t_{\rm 2,N}$ -- $t_{\rm 3,O}$ diagram
for H\,{\sc ii} regions  with $\log$ R$_3$ $<$ 0.5. 
The ff-based $t_{2,N}$ and $t_{3,O}$ temperatures in  
H\,{\sc ii} regions are used. 
The relation between electron temperature in low- and high-ionisation  
zones has been established on the basis of objects with weak R$_3$ line fluxes
$\log$ R$_3$ $<$ 0.5. The linear relation  
\begin{equation} 
t_{2,N} = 0.314 + 0.672 \, t_{3,O} . 
\label{equation:t2n-t3o}    
\end{equation} 
has been obtained by least-squares fitting. The coefficients in  
Eq.(\ref{equation:t2n-t3o}) were derived from the data for H\,{\sc ii} regions  
with weak R$_3$ line fluxes using an iterative fitting routine. Data points with large deviations were excluded in the fit.  
The resultant $t_{\rm 2,N}$ -- $t_{\rm 3,O}$ relation is shown by the solid line in  
Fig.~\ref{figure:t3o-t2n}.  
There is a trace of non-linearity in the $t_{\rm 2,N}$ -- $t_{\rm 3,O}$ 
diagram. The origin of this ``bending'' (if real) can be twofold:  
It can be artificial and caused by the fact that the particular form of the 
analytical expression adopted for the ff relation for [O\,{\sc iii}] lines 
and/or the ff relation for [N\,{\sc ii}]. We have chosen 
a simple form, but perhaps a more complex expression may give a better fit to the 
ff relation(s). It may also be that the linear expression adopted here for the 
$t_{\rm 2,N}$ -- $t_{\rm 3,O}$ relation is inappropriate. It is possible that the expression 
suggested by \citet{pagel1992} is more realistic. If so, the  
relations derived here may be considered as the first-order approximation. 
More high-precision measurements of nitrogen and oxygen  
auroral lines are needed to find better relations and to elucidate the  
origin of the possible non-linearity. 

The $t_{\rm 2,O}$ -- $t_{\rm 3,O}$ diagram 
(Fig.~\ref{figure:t3o-t2o}) has been examined in a similar way, and we thus obtain the expression
\begin{equation} 
t_{2,O} = 0.264 + 0.835 \,t_{3,O}.   
\label{equation:t2o-t3o}    
\end{equation} 
This $t_{\rm 2,O}$ -- $t_{\rm 3,O}$ relation is shown by the solid line in  
Fig.~\ref{figure:t3o-t2o}.  
The general properties of the $t_{\rm 2,O}$ -- $t_{\rm 3,O}$ diagram and  
the $t_{\rm 2,N}$ -- $t_{\rm 3,O}$ diagram are similar.  
The scatter in the $t_{\rm 2,N}$ -- $t_{\rm 3,O}$ diagram 
(Fig.~\ref{figure:t3o-t2n}) is smaller than that in the $t_{\rm 2,O}$ -- 
$t_{\rm 3,O}$ diagram (Fig.~\ref{figure:t3o-t2o}). This may suggest that 
the $t_{\rm 2,N}$ temperature is more reliable than the $t_{\rm 2,O}$ 
temperature.

\begin{figure} 
\resizebox{1.00\hsize}{!}{\includegraphics[angle=000]{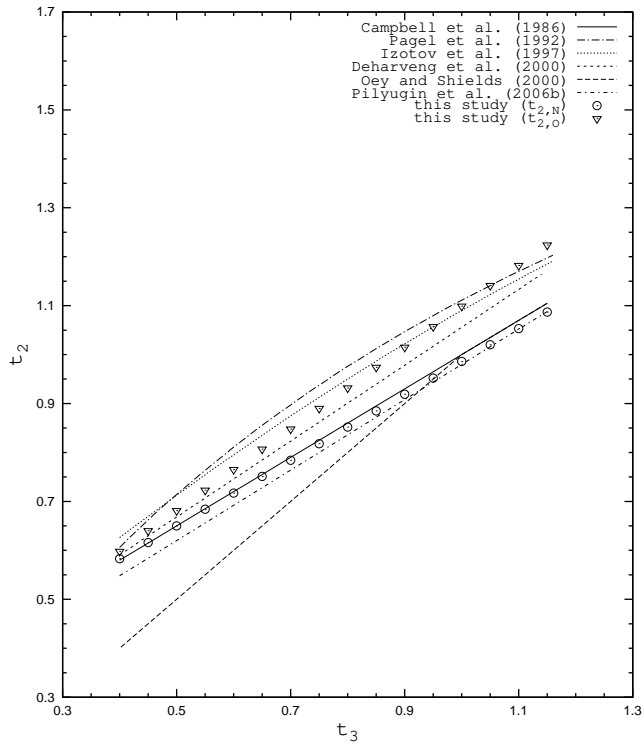}} 
\caption{Comparison of the t$_2$ -- t$_3$ relation derived here with those  
derived by other authors. The different lines are $t_2$ -- $t_3$ relations derived  
by these authors. Our $t_{2,N}$ -- $t_{3,O}$ relation is  
shown by the circles, and our $t_{2,O}$ -- $t_{3,O}$ relation is  
shown by the triangles. } 
\label{figure:t2comp} 
\end{figure}

\section{Discussion} 

Since the electron temperature $t_{\rm 2,N}$ seem to be more reliable  
than $t_{\rm 2,O}$, the $t_{\rm 2,N}$--$t_{\rm 3,O}$ diagram needs some 
further discussion.

\subsection{The comparison with previous t$_2$ -- t$_3$ relations} 

We compared the $t_{2,N}$ -- $t_{3,O}$ and $t_{2,O}$ -- $t_{3,O}$  
relations derived here with those obtained  
by other authors. Since our $t_2$ -- $t_3$ relations are derived for cool,  
high-metallicity H\,{\sc ii} regions, the high-temperature and low-metallicity  
part of the relation is not considered here. The different lines in Fig.~\ref{figure:t2comp}  
are the $t_2$ -- $t_3$ relations by  
\citet{campbell1986,pagel1992,izotov1997,deharveng2000,oey2000,pvt06}.  
Our $t_{2,N}$ -- $t_{3,O}$ and $t_{2,O}$ -- $t_{3,O}$ relations are shown by the open circles and  
the open triangles, respectively. Examination of Fig.~\ref{figure:t2comp} shows that  
our $t_{2,N}$ -- $t_{3,O}$ relation is very similar to the widely used  
$t_2$ -- $t_3$ relation after \citet{campbell1986} (see also \citet{garnett1992}).  

Just recently \citet{estebanetal09} derived a $t_{2,N}$ -- $t_{3,O}$ 
relation, which is very close to the one we have obtained here.

\begin{figure} 
\resizebox{1.00\hsize}{!}{\includegraphics[angle=000]{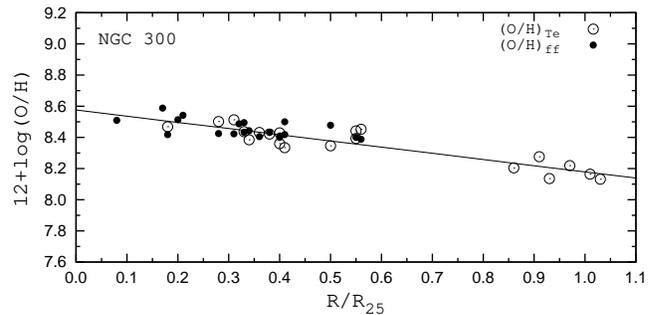}} 
\caption{The radial distribution of oxygen abundance in the disk of spiral 
galaxy NGC 300 traced by H\,{\sc ii} regions. The open circles are 
T$_{\rm e}$-based abundances, the line is the best fit to 
those data. The filled circles are ff-based abundances. The 
galactocentric distance is expressed in terms of the isophotal radius R$_{25}$. 
} 
\label{figure:ngc300} 
\end{figure} 

\subsection{Abundances in spiral galaxy NGC 300} 

New spectrophotometric 
observations of H\,{\sc ii} regions in the spiral galaxy NGC 300 have recently been 
published by \citet{bresolinetal09}. They have measured the auroral lines, and 
derived the radial oxygen abundance gradient in the disc of NGC 300 
based on ``direct'' (T$_{\rm e}$-based) oxygen abundances. This provides an additional 
possibility to test the relations we have derived in the present paper. 

T$_{\rm e}$-based oxygen abundances in H\,{\sc ii} regions, where the 
auroral oxygen line [O\,{\sc iii}]$\lambda$4363 is available, are shown in 
Fig.~\ref{figure:ngc300} by the open circles. The line is the linear best 
fit to those data. 
We have estimated the t$_3$ temperature from the ff relation, 
Eq.(\ref{equation:q3o}), and then the t$_2$ temperature from the t$_2$--t$_3$ 
relation, Eq.(\ref{equation:t2n-t3o}). 
Fig.~\ref{figure:ngc300} also shows that at the galactocentric distances larger than 
$\sim$ 0.7R$_{25}$, the oxygen abundance is lower than 12+log(O/H) = 8.2 (the 
R$_{25}$ is the isophotal radius). Therefore, the derived ff relation, 
Eq.~(\ref{equation:q3o}), is applicable only to the H\,{\sc ii} regions 
with galactocentric distances less than $\sim$ 0.7R$_{25}$. The derived 
(O/H)$_{\rm ff}$ abundances for 20 out of 21 H\,{\sc ii} regions are shown 
in Fig.~\ref{figure:ngc300} by the filled circles 
(the H\,{\sc ii} region \# 12 shows a very large deviation and was rejected). 
The (O/H)$_{\rm ff}$ abundances follow well the radial gradient traced by 
the (O/H)$_{\rm T_e}$ abundances. 
This confirms that the relations derived here,
result in realistic oxygen abundances.

\subsection{Deviations from the $t_2-t_3$ relation}  
 
\begin{figure} 
\resizebox{1.00\hsize}{!}{\includegraphics[angle=000]{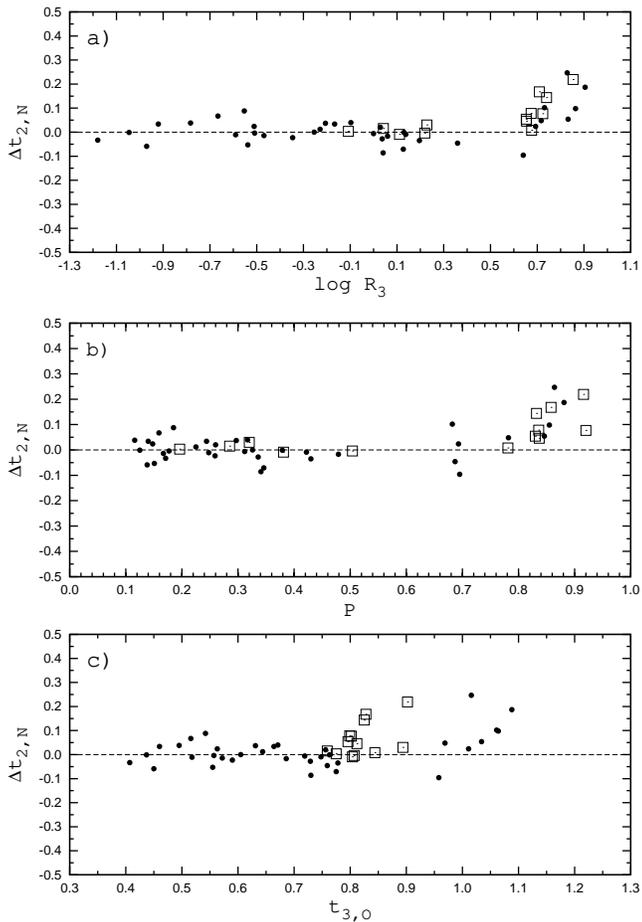}} 
\caption{ 
Deviations of $t_{\rm 2,N}$ from the $t_{\rm 2,N}$--$t_{\rm 3,O}$ relation as a 
function of R$_3$ line flux ({\it top panel}), excitation parameter ({\it middle 
panel}) 
and electron temperature $t_{\rm 3,O}$ ({\it bottom panel}). The filled circles 
show the deviations estimated using 
the measured $t_{2,N}$ and ff-based $t_{3,O}$ temperatures in H\,{\sc ii} 
regions from our sample.  The open squares 
show the deviations obtained using the measured $t_{3,O}$ and $t_{2,N}$ 
temperatures in galactic H\,{\sc ii} regions. 
} 
\label{figure:dt2n} 
\end{figure} 
 
We have found that the calibration data with strong R$_3$ lines  
($\log$~R$_3 \ga 0.5$) do not follow the ff relation for 
$Q_{\rm 2,N}$. Here, the deviations from the $t_{\rm 2,N}$ -- $t_{\rm 3,O}$ 
relation are examined. 
We define the deviation of $t_{\rm 2,N}$ from the $t_{\rm 2,N}$--$t_{\rm 3,O}$  
relation as the difference $\Delta$$t_{\rm 2,N}$ between the measured 
electron temperature $t_{\rm 2,N}$ and electron temperature  
t$^r_{\rm 2,N}$ obtained from the $t_{\rm 2,N}$--$t_{\rm 3,O}$ relation  
and the electron temperature $t_{\rm 3,O}$.  
Fig.~\ref{figure:dt2n} shows the deviations $\Delta$$t_{\rm 2,N}$ =  
$t_{\rm 2,N}$ -- t$^r_{\rm 2,N}$ as a function of  
R$_3$ line flux (top panel), excitation parameter $P$ (middle panel) 
and the electron temperature $t_{\rm 3,O}$ (bottom panel).  
Fig.~\ref{figure:t3o-t2n} shows that the H\,{\sc ii} regions with strong 
R$_3$ line fluxes show large deviations from the  
$t_{\rm 2,N}$--$t_{\rm 3,O}$ relation traced by the H\,{\sc ii} regions  
with weak R$_3$ line fluxes. 

High-precision spectroscopy, including the auroral line 
[N\,{\sc ii}]$\lambda$5755, for a number of galactic H\,{\sc ii} regions can 
be found in the literature \citep{estebanetal99a,estebanetal99b,   
estebanetal04,baldwinetal00,tsamisetal03,garciaetal04,garciaetal05,  
garciaetal06,garciaetal07}.   
These spectroscopic measurements cannot be used to construct the ff relations 
since only a small part of the galactic H\,{\sc ii} regions are measured and, 
therefore, the obtained excitation-parameter value is not representative for 
the whole nebula. However, they provide a remarkable possibility to test the  
derived $t_{\rm 2,N}$ -- $t_{\rm 2,O}$ relation. The open squares in 
Fig.~\ref{figure:dt2n} show the deviations $\Delta$$t_{\rm 2,N}$ of galactic 
H\,{\sc ii} regions.  The measured $t_{3,O}$ and $t_{2,N}$ temperatures  
in galactic H\,{\sc ii} regions are used. Inspection of Fig.~\ref{figure:dt2n} 
shows that the general behaviour of the deviations in the case of galactic 
H\,{\sc ii} regions is quite similar to that of extragalactic H\,{\sc ii} 
regions. The galactic H\,{\sc ii} regions with weak R$_3$ line fluxes follow the 
derived the $t_{\rm 2,N}$--$t_{\rm 3,O}$ relation 
(the deviations $\Delta$$t_{\rm 2,N}$ are small) and significant deviations 
occur in the case of galactic H\,{\sc ii} regions with strong R$_3$ line fluxes. 
Thus, the consideration of galactic H\,{\sc ii} regions confirms the results 
derived based on extragalactic H\,{\sc ii} regions. 

Fig.~\ref{figure:t3o-t2n} shows that there is a  
one-to-one correspondence between $t_{\rm 2,N}$ and  
$t_{\rm 3,O}$ within the uncertainties for H\,{\sc ii} regions with weak R$_3$  
lines. The H\,{\sc ii} regions with strong R$_3$ line fluxes do not follow the  
$t_{\rm 2,N}$--$t_{\rm 3,O}$ relation obtained from the H\,{\sc ii} regions  
with weak R$_3$ line fluxes, and the deviations $\Delta$$t_{\rm 2,N}$ do not depend 
on the electron temperature $t_{\rm 3,O}$.   
As a result, the one-to-one correspondence between electron temperatures  
$t_{\rm 2,N}$ and $t_{\rm 3,O}$ disappears if a sample contains  
H\,{\sc ii} regions with both weak and strong R$_3$ line fluxes.

\begin{figure} 
\resizebox{1.00\hsize}{!}{\includegraphics[angle=000]{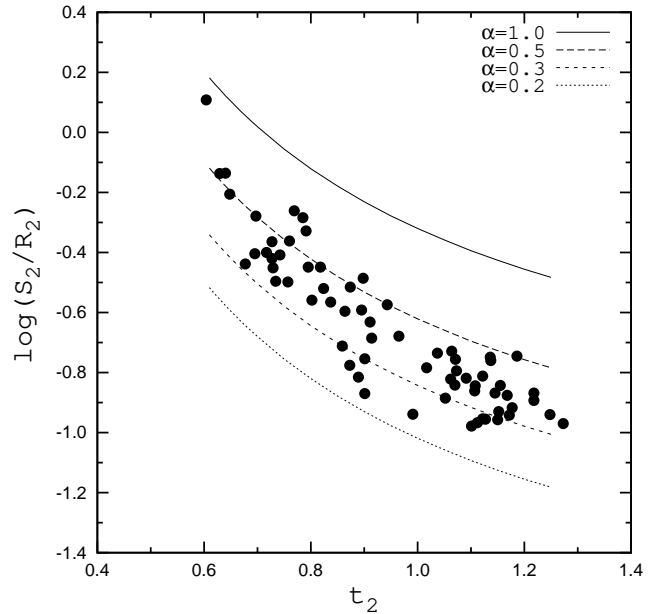}} 
\caption{ 
The S$_2$/R$_2$ line ratio as a function of electron temperature t$_2$. Family 
of theoretical curves for different values of the parameter $\alpha$ (which is 
the size ratio of S$^+$ and O$^+$ zones) superimposed on the observational data.
} 
\label{figure:so} 
\end{figure} 
 
\begin{figure} 
\resizebox{1.00\hsize}{!}{\includegraphics[angle=000]{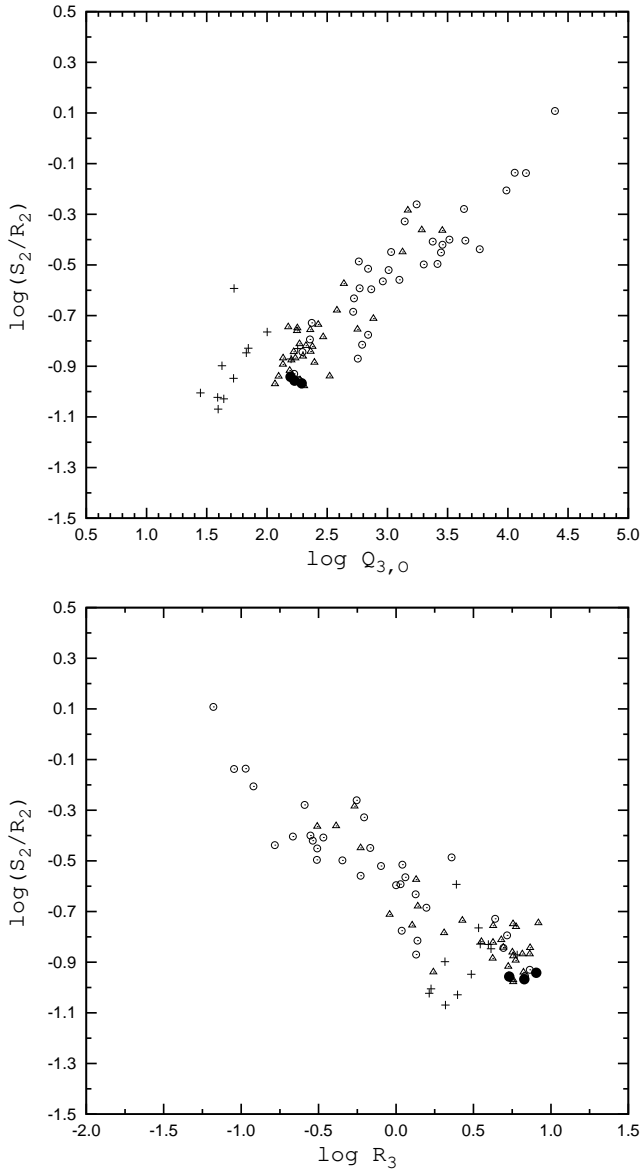}} 
\caption{ 
The S$_2$/R$_2$ line ratio as a function of Q$_{\rm 3,O}$ ({\it top panel}) 
and R$_3$ line flux ({\it bottom panel}). 
The open circles are high-metallicity H\,{\sc ii} regions with  
$\Delta$$t_{\rm 2,N}< 0.1$, the filled circles are those 
with $\Delta$$t_{\rm 2,N}> 0.1$, the open triangles are objects without 
$t_{\rm 2,N}$ measurements. 
The plus signs show the low-metallicity H\,{\sc ii} regions in the dwarf irregular 
galaxy DDO 68.
} 
\label{figure:min} 
\end{figure} 
 
\begin{figure} 
\resizebox{1.00\hsize}{!}{\includegraphics[angle=000]{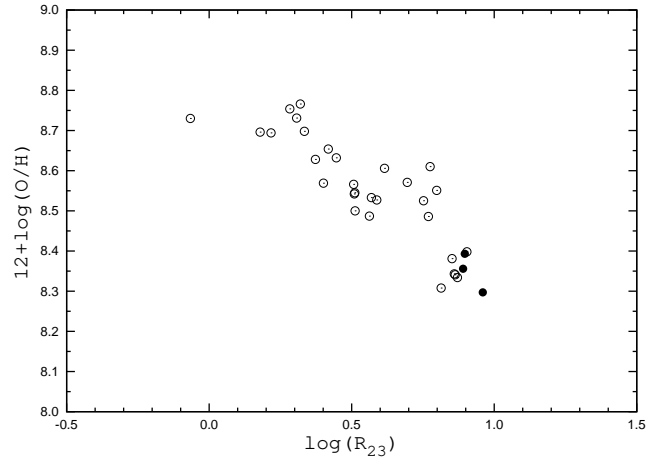}} 
\caption{ 
R$_{23}$ -- O/H diagram. The open circles are H\,{\sc ii} 
regions with $\Delta$$t_{\rm 2,N} < 0.1$ or with $\Delta$$t_{\rm 2,O} < 0.1$,  
the filled circles are those with $\Delta$$t_{\rm 2,N} > 0.1$.
} 
\label{figure:r23oh} 
\end{figure}

\subsection{A possible origin of the deviations from the $t_2-t_3$ relation} 

One may assume that the S/O abundance ratio is constant in all the H\,{\sc ii} 
regions since the sulfur and oxygen are thought to be produced by the same 
massive stars. It is commonly accepted that the  [O\,{\sc ii}] and [S\,{\sc ii}] 
lines are formed at similar temperatures. The energy of the level that gives 
rise to the R$_2$ line (the [O\,{\sc ii}]$\lambda$3727,3729 doublet) is higher than 
the energy of the level giving rise to the  
S$_2$ line (the [S\,{\sc ii}]$\lambda$6717,6731 doublet). Thus, the line-emissivity ratio 
$j$(S$_2$)/$j$(R$_2$) is temperature sensitive, i.e., the ratio increases with 
decreasing electron temperature. Furthermore, the S$^+$-zone does not 
coincide with the O$^+$-zone. Hence, the S$_2$/R$_2$ line ratio depends 
on two parameters; the electron temperature $t_2$ and the size ratio of S$^+$ 
and O$^+$ zones $\alpha$.  An analysis of the variation of the S$_2$/O$_2$ 
ratios in H\,{\sc ii} regions can help to illuminate the reason for the 
irregularity of high-R$_3$ H\,{\sc ii} regions. 

Fig.~\ref{figure:so} shows the measured S$_2$/R$_2$ ratio as a function of 
electron temperature $t_2$ (the values $t^r_{\rm 2,O}$ are used) for 
H\,{\sc ii} regions from our sample together with a family of theoretical 
curves computed for different values of $\alpha$ 
\begin{equation} 
\frac{S_2}{R_2} =  \alpha \, \frac{S}{O} \,\frac{j(S_2)}{j(R_2)} .
\label{equation:so}    
\end{equation} 
The solar S/O abundance ratio was used in the computations, i.e., $\log({\rm S/O}) = -1.50$ 
\citep{lodders2003}. The line emissivities $j$(S$_2$) and $j$(R$_2$) 
were derived from the five-level-atom solution for S$^+$ and  O$^+$.   
The energy levels and the Einstein coefficients for spontaneous transitions 
for five low-lying levels for S$^+$ were taken from \citet{irimia2003}, 
the effective collision strengths for the electron impact were taken from 
\citet{keenan1996}. The atomic data for O$^+$ were discussed above.  
Fig.~\ref{figure:so} shows a clear correlation between S$_2$/R$_2$ and $t_2$. 

Since there is a one-to-one correspondence between the $t_2$ and $t_3$ electron 
temperatures for high-metallicity H\,{\sc ii} regions, one might expect 
that there is also a correlation between S$_2$/R$_2$ and t$_3$. 
The use of the S$_2$/R$_2$ vs. $t_3$ instead of the S$_2$/R$_2$ vs. t$_2$ 
diagram has the advantage that the direct measurements of $t_{\rm 3,O}$ in 
high-R$_3$ H\,{\sc ii} regions are available in greater numbers than are direct 
measurements of $t_{\rm 2,O}$ and/or $t_{\rm 2,N}$. 
The top panel of Fig.~\ref{figure:min} shows the Q$_{\rm 3,O}$ vs. S$_2$/R$_2$ 
diagram. The open circles are H\,{\sc ii} regions with  
$\Delta$$t_{\rm 2,N}< 0.1$, the filled circles are those 
with $\Delta$$t_{\rm 2,N}> 0.1$, the open triangles are objects where 
$t_{\rm 2,N}$ is not measured. The low-metallicity 
(12+log(O/H) $\sim$ 7.2) H\,{\sc ii} regions in the dwarf galaxy DDO 68 
\citep{pustilnik2005} are shown for comparison (plus signs). 
Fig.~\ref{figure:min} (the top panel) shows a clear correlation between 
S$_2$/R$_2$ and Q$_{\rm 3,O}$; the S$_2$/R$_2$ line ratio decreases with 
decreasing of Q$_{\rm 3,O}$. 
In fact, the S$_2$/R$_2$ value can be used as an alternative indicator of the 
$t_3$, although the scatter in the Q$_{\rm 3,O}$ vs. S$_2$/R$_2$ relation is 
large. Fig.~\ref{figure:min} (top panel) shows that the H\,{\sc ii} regions 
with large deviations follow the general trend, although they are located near 
the end of the sequence. 

The bottom panel of Fig.~\ref{figure:min} shows a R$_{3}$ vs. S$_2$/R$_2$ 
diagram. The H\,{\sc ii} regions with $\log$~R$_3 > 0.5$ show a shift 
relative to the general trend obtained from the H\,{\sc ii} regions with 
$\log$~R$_3 < 0.5$. There are two possible causes for this shift. 
Either the high-R$_3$ and the low-R$_3$ H\,{\sc ii} regions do not belong to 
the same sequence (the upper branch), or 
the line fluxes of low-ionisation ions (N$^+$, O$^+$, S$^+$) are 
distorted in H\,{\sc ii} regions with strong R$_3$ line fluxes. 

Our sample only contains H\,{\sc ii} regions with 12+log(O/H) $>$ 
8.2. This selection criterion comes from the following consideration. 
It is well known that the relation between the oxygen abundance and the strong 
oxygen-line intensities is double valued with two distinct sequences, traditionally 
known as the upper and the lower branches of the R$_{23}$ -- O/H diagram. 
The transition zone between the upper and the lower branches is sometimes also taken
into consideration \citep{pilyuginthuan05}.    
We adopt 12+log(O/H) = 8.2 as the boundary between the upper branch and the 
transition zone. This division is somewhat arbitrary, however. 
Fig.~\ref{figure:r23oh} shows the R$_{23}$ -- O/H diagram for 
the H\,{\sc ii} regions with measured 
$t_{\rm 2,N}$ or $t_{\rm 2,O}$ temperature. 
The question is, if one should adopt a higher oxygen abundance (e.g., 12+log(O/H) = 8.4) as 
the boundary between the upper branch and the transition zone?
We believe that is probably not the way to solve the problem. 
It should be noted that both H\,{\sc ii} regions with small deviations and 
H\,{\sc ii} regions with large deviations follow the ff relation for 
$Q_{\rm 3,O}$ rather well. Furthermore,  
Fig.~\ref{figure:r23oh} shows that the area occupied by the H\,{\sc ii} regions 
with large deviations is also populated by the H\,{\sc ii} regions with small 
deviations. 
Finally, some high-metallicity galactic H\,{\sc ii} regions also show large 
deviations. 
It is possible that a criterion other than a specific metallicity should be used to divide 
the H\,{\sc ii} regions into sequences (branches).

The deviations suggest that the $Q_{\rm 2,N}$ values are shifted in  
H\,{\sc ii} regions with strong R$_3$ line fluxes (high excitation parameter $P$).  
It has been suggested that the low-lying metastable levels in some ions are  
excited not only by the electron collisions, but may also be significantly  
populated by recombination processes \citep{rubin86,liuetal00}. Since  
recombination-excited emission of [O\,{\sc ii}] in the auroral  
$\lambda$7325, and [N\,{\sc ii}] in the auroral $\lambda$5755 radiation will  
occur in the O$^{++}$ zone, the effect on the  
collision-excited lines will depend on the R$_3$ line flux. As  
a result, one may expect a significant shift of the $Q_{\rm 2,N}$ and  
$Q_{\rm 2,O}$ values for H\,{\sc ii} regions with strong R$_3$ line fluxes. Our  
results are generally in agreement with this picture. However, other mechanisms 
causing a shift of this kind of the $Q_{\rm 2,N}$ value cannot be excluded.

\section{Conclusions}  
 
The electron temperatures of high-metallicity (12+log(O/H) $>$ 8.2) H\,{\sc ii} regions have been studied.  
Empirical ff relations which express the nebular-to-auroral lines ratios  
($Q_{\rm 3,O}$, $Q_{\rm 2,O}$, and $Q_{\rm 2,N}$) in terms of the nebular  
R$_3$ and R$_2$ line fluxes in spectra of H\,{\sc ii} regions   
have been derived. The ff relations for $Q_{\rm 2,O}$ and $Q_{\rm 2,N}$ only reproduce the 
observational data in spectra of H\,{\sc ii} regions with weak nebular  
R$_3$ line fluxes (logR$_3$ $\la$ 0.5). The ff relation for $Q_{\rm 3,O}$, on the other hand, reproduces  
the observational data over the whole range of nebular R$_3$ line fluxes. 
 
The $t_2$ -- $t_3$ diagram has also been considered. We found that there is a
one-to-one correspondence between the $t_2$ and $t_3$ electron temperatures, within the 
uncertainties of these electron temperatures, for H\,{\sc ii} regions with   
weak nebular R$_3$ line fluxes. The derived $t_{2,N}$ -- $t_{3,O}$ relation  
for these H\,{\sc ii} regions is very similar to the commonly used $t_2$ -- $t_3$  
relation after \citet{campbell1986}. The H\,{\sc ii} regions with strong  
nebular R$_3$ line fluxes do not follow this relation, however. Thus, as a result, the  
one-to-one correspondence between the electron temperatures $t_{\rm 2,N}$ and  
$t_{\rm 3,O}$ disappears if H\,{\sc ii} regions with both weak and strong R$_3$ line fluxes are included.  
 
There is a discrepancy between $t_{\rm 2,N}$ and $t_{\rm 2,O}$ temperatures.  
The $t_{\rm 2,N}$ temperatures are systematically lower than the $t_{\rm 2,O}$  
temperatures. The difference is small at low electron temperatures and  
increases with the electron temperature to about  
10\% at $t = 1$. The uncertainties in the atomic data (Einstein coefficients for  
spontaneous transitions and effective cross sections for electron impact) may  
be the cause of this discrepancy. 

\section*{Acknowledgments}

We thank the referee, F.P.~Keenan, for constructive criticism and useful comments that helped to improve the paper. We also thank N. Bergvall for
useful discussions on line-blending effects.

\appendix

\section{A sample of the calibration H\,{\sc ii} regions. Online material}

Table~\ref{table:lines} contains the line intensities measured in the spectra 
of the calibration H\,{\sc ii} regions.
The first column gives the object name (which is typically the name 
of the parent galaxy and the H\,{\sc ii} region identifier). The second column 
provides the data-source reference. 
The following abbreviations for the references are used:\\
G00 -- \citet{guseva00}, \\ 
L02 -- \citet{luridianaetal02}, \\ 
V02 -- \citet{vermeijetal02}, \\  
K03 -- \citet{kennicuttetal03}, \\ 
L03 -- \citet{leeetal03}, \\ 
T03 -- \citet{tsamisetal03},  \\
P03 -- \citet{peimbert03}, \\ 
V03 -- \citet{vilchez03}, \\ 
B04 -- \citet{bresolinetal04}, \\ 
I04 -- \citet{izotovthuan04}, \\ 
G04 -- \citet{garnettetal04}, \\ 
L04a -- \citet{leeetal04}, \\ 
L04b -- \citet{leeskillman04}, \\ 
B05 -- \citet{bresolinetal05}, \\ 
T05 -- \citet{thuan05}, \\  
C06 -- \citet{crockett06}, \\
B07 -- \citet{bresolin07}.  \\
The measured 
[O\,{\sc ii}]$\lambda$3727+$\lambda$3729, 
[O\,{\sc iii}]$\lambda$4363, 
[O\,{\sc iii}]$\lambda$4959+$\lambda$5007, 
[N\,{\sc ii}]$\lambda$5755, 
[N\,{\sc ii}]$\lambda$6548+$\lambda$6584, 
[S\,{\sc ii}]$\lambda$6717+$\lambda$6731,
[O\,{\sc ii}]$\lambda$7320+$\lambda$7330 
line intensities are listed in columns from 3 to 9 respectively.
The line intensities are given on a scale where H$\beta=1$.


\setcounter{table}{0}
\begin{table*}
\caption[]{\label{table:lines}
Line intensities of the sample of calibration H\,{\sc ii} regions. 
The line intensities are given on a scale where H$\beta=1$.
}
\begin{center}
\begin{tabular}{lcccccccc} \hline \hline
Object name                                              & 
Reference                                                & 
[O\,{\sc ii}]                                            & 
[O\,{\sc iii}]                                           & 
[O\,{\sc iii}]                                           & 
[N\,{\sc ii}]                                            & 
[N\,{\sc ii}]                                            & 
[S\,{\sc ii}]                                            & 
[O\,{\sc ii}]                                            \\
identificator                                            & 
                                                         & 
$\lambda$3727                                            & 
$\lambda$4363                                            & 
$\lambda$4959+5007                                       & 
$\lambda$5755                                            & 
$\lambda$6548+6584                                       & 
$\lambda$6717+6731                                       & 
$\lambda$7320+7330                                       \\   \hline
 Mrk 1329     & G00  &   1.152 &   0.0460 &   7.308 &   0.00000 &   0.099 &   0.156 &   0.0000 \\ 
 M101 N5461   & L02  &   1.914 &   0.0119 &   4.361 &   0.00400 &   0.566 &   0.357 &   0.0415 \\ 
 SMC  N160A1  & V02  &   1.655 &   0.0000 &   5.635 &   0.00000 &   0.249 &   0.228 &   0.0460 \\ 
 LMC  N4A     & V02  &   1.522 &   0.0000 &   5.720 &   0.00000 &   0.177 &   0.160 &   0.0400 \\ 
 M101 H67     & K03  &   2.440 &   0.0350 &   4.560 &   0.00000 &   0.217 &   0.263 &   0.0540 \\ 
 M101 H70     & K03  &   3.110 &   0.0250 &   3.560 &   0.00000 &   0.315 &   0.472 &   0.0000 \\ 
 M101 H128    & K03  &   1.450 &   0.0170 &   5.213 &   0.00300 &   0.277 &   0.233 &   0.0410 \\ 
 M101 H149    & K03  &   2.120 &   0.0180 &   4.240 &   0.00000 &   0.479 &   0.372 &   0.0560 \\ 
 M101 H336    & K03  &   1.780 &   0.0000 &   0.308 &   0.00500 &   1.278 &   0.568 &   0.0140 \\ 
 M101 H409    & K03  &   2.180 &   0.0230 &   4.931 &   0.00400 &   0.364 &   0.312 &   0.0470 \\ 
 M101 H1013   & K03  &   1.880 &   0.0000 &   1.373 &   0.00500 &   0.861 &   0.288 &   0.0260 \\ 
 M101 H1105   & K03  &   1.850 &   0.0140 &   4.210 &   0.00000 &   0.445 &   0.241 &   0.0360 \\ 
 M101 H1159   & K03  &   1.980 &   0.0190 &   4.227 &   0.00000 &   0.315 &   0.298 &   0.0480 \\ 
 M101 H1170   & K03  &   3.080 &   0.0160 &   2.680 &   0.00000 &   0.587 &   0.567 &   0.0000 \\ 
 M101 H1176   & K03  &   1.600 &   0.0240 &   4.920 &   0.00000 &   0.283 &   0.230 &   0.0300 \\ 
 IC10 \# 1    & L03  &   1.980 &   0.0237 &   4.834 &   0.00000 &   0.000 &   0.000 &   0.0000 \\ 
 IC10 \# 2    & L03  &   1.750 &   0.0220 &   4.929 &   0.00000 &   0.000 &   0.000 &   0.0000 \\ 
 LMC 30 Dor   & T03  &   1.056 &   0.0335 &   6.730 &   0.00206 &   0.120 &   0.114 &   0.0377 \\ 
 LMC 30 Dor   & P03  &   1.235 &   0.0321 &   6.788 &   0.00187 &   0.152 &   0.137 &   0.0373 \\ 
 VCC1699      & V03  &   1.260 &   0.0410 &   7.350 &   0.00000 &   0.104 &   0.181 &   0.0000 \\ 
 VCC144       & V03  &   2.541 &   0.0240 &   4.782 &   0.00000 &   0.274 &   0.392 &   0.0000 \\ 
 M51 CCM10    & B04  &   1.260 &   0.0000 &   0.165 &   0.00500 &   1.490 &   0.460 &   0.0000 \\ 
 M51 CCM53    & B04  &   1.290 &   0.0000 &   0.451 &   0.00540 &   1.600 &   0.410 &   0.0000 \\ 
 M51 CCM54    & B04  &   1.150 &   0.0000 &   0.557 &   0.00650 &   1.620 &   0.630 &   0.0000 \\ 
 M51 CCM55    & B04  &   0.780 &   0.0000 &   0.257 &   0.00430 &   1.550 &   0.410 &   0.0000 \\ 
 M51 CCM57    & B04  &   1.140 &   0.0000 &   0.216 &   0.00680 &   1.640 &   0.450 &   0.0000 \\ 
 M51 CCM57A   & B04  &   1.040 &   0.0000 &   2.286 &   0.00480 &   0.930 &   0.340 &   0.0000 \\ 
 M51 CCM71A   & B04  &   1.470 &   0.0000 &   0.622 &   0.00790 &   1.550 &   0.690 &   0.0000 \\ 
 M51 CCM72    & B04  &   0.630 &   0.0000 &   0.090 &   0.00280 &   1.330 &   0.460 &   0.0000 \\ 
 M51 CCM84A   & B04  &   1.250 &   0.0000 &   1.147 &   0.00920 &   1.930 &   0.340 &   0.0000 \\ 
 M51 P203     & B04  &   0.320 &   0.0000 &   0.066 &   0.00150 &   1.050 &   0.410 &   0.0000 \\ 
 HS0111+2115  & I04  &   3.207 &   0.0389 &   5.928 &   0.00000 &   0.259 &   0.410 &   0.0000 \\ 
 UM396        & I04  &   1.117 &   0.0588 &   8.290 &   0.00000 &   0.108 &   0.201 &   0.0000 \\ 
 Mrk 35       & I04  &   2.509 &   0.0273 &   5.380 &   0.00370 &   0.265 &   0.277 &   0.0468 \\ 
 Mrk 1315     & I04  &   1.085 &   0.0522 &   8.039 &   0.00130 &   0.078 &   0.124 &   0.0000 \\ 
 Mrk 1329     & I04  &   1.242 &   0.0443 &   7.309 &   0.00130 &   0.094 &   0.146 &   0.0208 \\ 
 M51 CCM72    & I04  &   0.670 &   0.0000 &   0.107 &   0.00200 &   1.380 &   0.490 &   0.0000 \\ 
 KISSR 87     & L04a &   1.895 &   0.0280 &   5.958 &   0.00000 &   0.302 &   0.329 &   0.0000 \\ 
 KISSR 286    & L04a &   2.053 &   0.0370 &   5.683 &   0.00000 &   0.204 &   0.366 &   0.0400 \\ 
 NGC1705 A3   & L04b &   2.540 &   0.0410 &   5.720 &   0.00000 &   0.124 &   0.338 &   0.0000 \\ 
 NGC1705 B3   & L04b &   2.740 &   0.0560 &   6.530 &   0.00000 &   0.131 &   0.371 &   0.0000 \\ 
 NGC1705 B4   & L04b &   3.120 &   0.0480 &   6.650 &   0.00000 &   0.092 &   0.358 &   0.0000 \\ 
 NGC1705 B6   & L04b &   4.360 &   0.0570 &   5.720 &   0.00000 &   0.197 &   0.467 &   0.0000 \\ 
 NGC1705 C6   & L04b &   3.018 &   0.0330 &   5.303 &   0.00000 &   0.083 &   0.365 &   0.0000 \\ 
 NGC1232 \# 2 & B05  &   3.910 &   0.0000 &   1.740 &   0.00000 &   0.640 &   0.450 &   0.0840 \\ 
 NGC1232 \# 4 & B05  &   2.070 &   0.0000 &   5.360 &   0.00000 &   0.290 &   0.230 &   0.0400 \\ 
 NGC1232 \# 5 & B05  &   2.530 &   0.0000 &   1.340 &   0.00540 &   1.110 &   0.590 &   0.0340 \\ 
 NGC1232 \# 6 & B05  &   2.110 &   0.0000 &   0.680 &   0.00660 &   1.200 &   0.750 &   0.0320 \\ 
 NGC1232 \# 7 & B05  &   1.630 &   0.0000 &   0.290 &   0.00390 &   1.560 &   0.620 &   0.0140 \\ 
 NGC1232 \# 10& B05  &   1.340 &   0.0000 &   0.310 &   0.00000 &   1.230 &   0.580 &   0.0120 \\ 
 NGC1232 \# 14& B05  &   3.050 &   0.0000 &   1.070 &   0.00840 &   1.270 &   0.780 &   0.0540 \\ 
 NGC1232 \# 15& B05  &   4.580 &   0.0000 &   1.380 &   0.00000 &   0.760 &   0.960 &   0.0910 \\ 
 NGC1365 \# 3 & B05  &   2.420 &   0.0000 &   0.910 &   0.00000 &   1.130 &   0.470 &   0.0240 \\ 
 NGC1365 \# 5 & B05  &   1.690 &   0.0000 &   0.340 &   0.00450 &   1.350 &   0.660 &   0.0160 \\ 
 NGC1365 \# 8 & B05  &   1.720 &   0.0000 &   0.800 &   0.00660 &   1.150 &   0.520 &   0.0200 \\ 
 NGC1365 \# 12& B05  &   1.750 &   0.0000 &   0.410 &   0.00000 &   1.380 &   0.760 &   0.0160 \\ 
 NGC1365 \# 15& B05  &   2.210 &   0.0000 &   1.000 &   0.00920 &   1.690 &   0.560 &   0.0310 \\ 
 NGC1365 \# 16& B05  &   2.440 &   0.0000 &   1.270 &   0.00000 &   0.910 &   0.430 &   0.0400 \\ 
 NGC1365 \# 17& B05  &   3.600 &   0.0000 &   1.350 &   0.00000 &   1.150 &   0.960 &   0.0560 \\ 
\hline 
\end{tabular}\\
\end{center}
\end{table*}


\setcounter{table}{0}
\begin{table*}
\caption[]{
Continued
}
\begin{center}
\begin{tabular}{lcccccccc} \hline \hline
Object name                                              & 
Reference                                                & 
[O\,{\sc ii}]                                            & 
[O\,{\sc iii}]                                           & 
[O\,{\sc iii}]                                           & 
[N\,{\sc ii}]                                            & 
[N\,{\sc ii}]                                            & 
[S\,{\sc ii}]                                            & 
[O\,{\sc ii}]                                            \\
                                                         & 
                                                         & 
$\lambda$3727                                            & 
$\lambda$4363                                            & 
$\lambda$4959+5007                                       & 
$\lambda$5755                                            & 
$\lambda$6548+6584                                       & 
$\lambda$6717+6731                                       & 
$\lambda$7320+7330                                       \\   \hline
 NGC2997 \# 4 & B05  &   3.830 &   0.0000 &   2.040 &   0.00000 &   0.800 &   0.630 &   0.0660 \\ 
 NGC2997 \# 5 & B05  &   2.130 &   0.0000 &   1.100 &   0.00460 &   1.160 &   0.650 &   0.0340 \\ 
 NGC2997 \# 6 & B05  &   2.150 &   0.0000 &   1.090 &   0.00580 &   1.130 &   0.360 &   0.0320 \\ 
 NGC2997 \# 7 & B05  &   2.030 &   0.0000 &   0.590 &   0.00640 &   1.340 &   0.560 &   0.0250 \\ 
 NGC2997 \# 11& B05  &   1.710 &   0.0000 &   0.540 &   0.00000 &   1.410 &   0.890 &   0.0270 \\ 
 NGC2997 \# 14& B05  &   1.770 &   0.0000 &   0.590 &   0.00000 &   1.290 &   0.630 &   0.0240 \\ 
 NGC5236 \# 3 & B05  &   2.080 &   0.0000 &   1.570 &   0.00740 &   1.300 &   0.430 &   0.0290 \\ 
 NGC5236 \# 6 & B05  &   1.230 &   0.0000 &   0.280 &   0.00810 &   1.650 &   0.490 &   0.0120 \\ 
 NGC5236 \# 14& B05  &   0.740 &   0.0000 &   0.120 &   0.00420 &   1.460 &   0.460 &   0.0060 \\ 
 NGC5236 \# 16& B05  &   1.440 &   0.0000 &   0.310 &   0.00600 &   1.790 &   0.510 &   0.0000 \\ 
 Mrk 35 \# 1  & T05  &   1.874 &   0.0265 &   5.408 &   0.00000 &   0.000 &   0.000 &   0.0000 \\ 
 M33 BCLMP090 & C06  &   1.980 &   0.0410 &   8.760 &   0.00000 &   0.000 &   0.000 &   0.0000 \\ 
 M33 BCLMP691 & C06  &   1.680 &   0.0220 &   4.560 &   0.00000 &   0.000 &   0.000 &   0.0000 \\ 
 M33 BCLMP706 & C06  &   2.790 &   0.0120 &   2.910 &   0.00000 &   0.000 &   0.000 &   0.0000 \\ 
 M33 BCLMP290a& C06  &   2.230 &   0.0110 &   2.490 &   0.00000 &   0.000 &   0.000 &   0.0000 \\ 
 M33 MA 1     & C06  &   1.140 &   0.0430 &   6.980 &   0.00000 &   0.000 &   0.000 &   0.0000 \\ 
 M101 H1013   & B07  &   2.210 &   0.0024 &   1.350 &   0.00590 &   0.950 &   0.298 &   0.0240 \\ 
\hline 
\end{tabular}\\
\end{center}
\end{table*}


\begin{thebibliography}{} 
\bibitem [Aggarwal \& Keenan (1999)]{aggarwal1999} 
          Aggarwal K.M., Keenan F.P. 1999, ApJS, 123, 31 
\bibitem [Baldwin et al. (2000)]{baldwinetal00} 
          Baldwin J.A., Verner E.M., Verner D.A., Ferland G.J., Martin P.G.,  
          Korista K.T., Rubin R.H., 2000, ApJS, 129, 229 
\bibitem [Bresolin et al. (2004)]{bresolinetal04} 
          Bresolin F., Garnett D.R., Kennicutt R.C., 2004, ApJ, 615, 228 
\bibitem [Bresolin et al. (2005)]{bresolinetal05} 
          Bresolin F., Schaerer D., Conz\'{a}lez Delgado R.M.,  
          Stasi\'{n}ska, G. 2005, A\&A, 441, 981 
\bibitem [Bresolin (2007)]{bresolin07} 
          Bresolin F., 2007, ApJ, 656, 186 
\bibitem [Bresolin et al. (2009)]{bresolinetal09} 
          Bresolin F., Gieren W., Kudritzki R.-P., Pietrzy\'{n}ski G., Urbaneja M.A., 
          2009, ApJ, accepted (astro-ph/0905.2791)
\bibitem [Campbell, Terlevich \& Melnick (1986)]{campbell1986}  
          Campbell A., Terlevich R., Melnick J., 1986, MNRAS, 223, 811 
\bibitem [Crockett et al (2006)]{crockett06} 
          Crockett N.R., Garnett D.R., Massey P., Jacoby G.  
          2006, ApJ, 637, 741 
\bibitem[Deharveng et al.(2000)]{deharveng2000}  
         Deharveng L., Pe\~{n}a M., Caplan J., Costero R.,   
         2000, MNRAS, 311, 329 
\bibitem [Edl\'en (1985)]{edlen1985} 
          Edl\'en B. 1985, Phys. Scripta, 31, 345 
\bibitem [Esteban et al. (1999a)]{estebanetal99a}  
          Esteban C., Peimbert M., Torres-Peimbert S., Garc\'{\i}a-Rojas J.,  
          1999a, Rev..Mex. A\&A, 35, 65 
\bibitem [Esteban et al. (1999b)]{estebanetal99b}  
          Esteban C., Peimbert M., Torres-Peimbert S., Garc\'{\i}a-Rojas J.,  
          Rodr\'{i}guez M., 1999b, ApJS, 120, 113 
\bibitem [Esteban et al. (2004)]{estebanetal04}  
          Esteban C., Peimbert M., Garc\'{\i}a-Rojas J., Ruiz M.T.,  
          Peimbert A., Rodr\'{i}guez M., 2004, MNRAS, 355, 229 
\bibitem [Esteban et al. (2009)]{estebanetal09}  
          Esteban C., Peimbert M., Garc\'{\i}a-Rojas J., Peimbert A., 
          Mesa-Delgado A., 2009, ApJ, accepted (astro-ph/0905.2532)
\bibitem [Froese Fisher \& Tachiev (2004)]{froese2004}  
          Froese Fischer C., Tachiev G. 2004, ADNDT, 87, 1 
\bibitem [Galav\'{i}s, Mendoza \& Zeippen (1997)]{galavis1997} 
          Galav\'{i}s M.E., Mendoza C., Zeippen C.J. 1997, A\&AS, 123, 159  
\bibitem [Garc\'{\i}a-Rojas et al. (2004)]{garciaetal04}  
          Garc\'{\i}a-Rojas J., Esteban C., Peimbert M., Rodr\'{i}guez M.,  
          Ruiz M.T., Peimbert A., 2004, ApJS, 153, 501 
\bibitem [Garc\'{\i}a-Rojas et al. (2005)]{garciaetal05}  
          Garc\'{\i}a-Rojas J., Esteban C., Peimbert A., Peimbert M.,  
          Rodr\'{i}guez M., Ruiz M.T., 2005, MNRAS, 362, 301 
\bibitem [Garc\'{\i}a-Rojas et al. (2006)]{garciaetal06}  
          Garc\'{\i}a-Rojas J., Esteban C., Peimbert M., Costado M.T.,  
          Rodr\'{i}guez M., Peimbert M., Ruiz M.T., 2006, MNRAS, 368, 253 
\bibitem [Garc\'{\i}a-Rojas et al. (2007)]{garciaetal07}  
          Garc\'{\i}a-Rojas J., Esteban C., Peimbert A., Rodr\'{i}guez M.,  
          Peimbert M., Ruiz M.T., 2007, Rev.Mex.A\&A, 43, 3 
\bibitem [Garnett (1992)]{garnett1992} 
          Garnett D.R., 1992, AJ, 103, 1330  
\bibitem [Garnett, Kennicutt \& Bresolin (2004)]{garnettetal04} 
          Garnett D., Kennicutt R.C., Bresolin F. 2004, ApJ, 607, L21  
\bibitem [Guseva, Izotov \& Thuan (2000)]{guseva00} 
          Guseva N.G., Izotov Y.I., Thuan T.X. 2000, ApJ, 531, 776  
\bibitem [Hudson \& Bell (2004)]{hudson2004} 
          Hudson C.E., Bell K.L. 2004,  MNRAS,  348, 1275
\bibitem [Hudson \& Bell (2005)]{hudson2005} 
          Hudson C.E., Bell K.L. 2005,  A\&A,  430, 725 
\bibitem[Irimia \& Froese Fisher (2003)]{irimia2003}
         Irimia A., Froese Fisher C., 2003, http://hf8.vuse.vanderbilt.edu 
\bibitem [Izotov \& Thuan (2004)]{izotovthuan04} 
          Izotov Y.I., Thuan T.X. 2004, ApJ, 602, 200  
\bibitem[Izotov, Thuan \& Lipovetsky (1997)]{izotov1997} 
         Izotov Y.I., Thuan T.X., Lipovetsky V.A., 1997, ApJS, 108, 1 
\bibitem[Keenan et al.(1996)]{keenan1996} 
         Keenan F.P., Aller L.H., Bell K.L., Hyung S., McKenna F.C., 
         Ramsbottom C.A., 1996, MNRAS 281, 1073
\bibitem [Kennicutt, Bresolin \& Garnett (2003)]{kennicuttetal03} 
          Kennicutt R.C., Bresolin F., Garnett D. 2003, ApJ, 591, 801  
\bibitem [Lee et al. (2003)]{leeetal03} 
          Lee H., McCall M.L., Kingsburch R.L., Ross R., Stevenson C.  
          2003, AJ, 125, 146 
\bibitem [Lee \& Skillman (2004)]{leeskillman04} 
          Lee H., Skillman E.D. 2004, ApJ, 614, 698 
\bibitem [Lee, Salzer \& Melbourne (2004)]{leeetal04} 
          Lee H., Salzer J.J., Melbourne J. 2004, ApJ, 616, 752 
\bibitem [Liu et al. (2000)]{liuetal00} 
          Liu X.-W., Storey P.J., Barlow M.J., Danziger I.J., Cohen M.,  \&  
          Bryce M., 2000, MNRAS, 312, 585  
\bibitem [Lodders (2003)]{lodders2003} 
          Lodders K. 2003, ApJ, 591, 1220
\bibitem [Luridiana et al. (2002)]{luridianaetal02} 
          Luridiana V., Esteban C., Peimbert M., Peimbert A.  
          2002, RevMexAA, 38, 97 
\bibitem[Oey \& Shields (2000)]{oey2000}  
         Oey M.S., Shields J.C., 2000, ApJ, 539, 687 
\bibitem [Pagel et al. (1992)]{pagel1992}  
          Pagel, B.E.J., Simonson, E.A., Terlevich, R.J., \& Edmunds, M.G.  
          1992, MNRAS, 255, 325 
\bibitem [Peimbert (2003)]{peimbert03}  
          Peimbert A., 2003, ApJ, 584, 735 
\bibitem [Pilyugin (2000)]{lcal} 
          Pilyugin L.S., 2000, A\&A, 362, 325 
\bibitem [Pilyugin (2001)]{hcal} 
          Pilyugin L.S., 2001, A\&A, 369, 594 
\bibitem [Pilyugin (2003)]{vybor} 
          Pilyugin L.S., 2003, A\&A, 399, 1003 
\bibitem [Pilyugin(2005)]{ff} 
          Pilyugin L.S., 2005, A\&A, 436, 1L 
\bibitem [Pilyugin (2007)]{pilyugin07} 
          Pilyugin L.S. 2007, MNRAS, 375, 685 
\bibitem [Pilyugin (2009)]{pilyugin09} 
          Pilyugin L.S. 2009, Kinematika i fisika nebesnyh tel (in russian),  
          accepted 
\bibitem [Pilyugin, Thuan \& V\'{i}lchez (2006)]{pilyuginetal06} 
          Pilyugin L.S., Thuan T.X., V\'{i}lchez J.M., 2006a, MNRAS, 367, 1139 
\bibitem [Pilyugin \& Thuan (2005)]{pilyuginthuan05}  
          Pilyugin L.S., Thuan T.X., 2005, ApJ, 631, 231 
\bibitem [Pilyugin \& Thuan (2007)]{pilyuginthuan07}  
          Pilyugin L.S., Thuan T.X., 2007, ApJ, 669, 290 
\bibitem [Pilyugin, V\'{i}lchez \& Thuan (2006)]{pvt06} 
          Pilyugin L.S., V\'{i}lchez J.M., Thuan T.X., 2006b, MNRAS, 370, 1928 
\bibitem [Pradhan et. al. (2006)]{pradhan2006}  
          Pradhan A.K., Montenegro M., Nahar S.N., Eissner W. 
          2006, MNRAS, 366, L6 
\bibitem [Pustilnik, Kniazev \& Pramskij (2005)]{pustilnik2005} 
          Pustilnik S.A., Kniazev A.Y., Pramskij A.G. 
          2005, A\&A, 443, 91
\bibitem [Rubin (1986)]{rubin86} 
          Rubin R.H., 1986, ApJ, 309, 334 
\bibitem [Thuan \& Izotov (2005)]{thuan05}  
          Thuan T.X., Izotov Y.I. 2005, ApJS, 161, 240  
\bibitem [Tayal (2007)]{tayal2007}  
          Tayal S.S., 2007, ApJS, 171, 331
\bibitem [Tsamis et al. (2003)]{tsamisetal03}  
          Tsamis Y.G., Barlow M.J., Liu X.-W., Danziger I.J., Storey P.J.,  
          2003, MNRAS, 338, 687 
\bibitem [Vermeij et al. (2002)]{vermeijetal02}  
          Vermeij R., Damour F., van der Hulst, J.M., Baluteau J.-P.,  
          2002, A\&2, 390, 649 
\bibitem [V\'{i}lchez \& Iglesias-P\'{a}ramo (2003)]{vilchez03} 
          V\'{i}lchez J.M., Iglesias-P\'{a}ramo J. 2003, ApJS, 145, 225 
\bibitem[Wen\aa ker (1990)]{wenaker1990} 
         Wen\aa ker I. 1990, Phys. Scripta, 42, 667  
\bibitem[Zaritsky, Kennicutt \& Huchra (1994)]{zkh} 
         Zaritsky D., Kennicutt R.C., Huchra J.P. 1994,  
         ApJ, 420, 87 
\end{thebibliography}
\end{document}